\documentclass[conference]{IEEEtran}

\usepackage[utf8]{inputenc}
\usepackage[T1]{fontenc}
\usepackage{microtype}

\usepackage{amsmath,amssymb,amsthm}
\usepackage{mathtools}

\usepackage{graphicx}
\usepackage{xcolor}
\definecolor{nvidiagreen}{HTML}{76B900}
\definecolor{pastelpurple}{HTML}{D8C7FF}
\definecolor{figink}{HTML}{23313F}
\definecolor{figamber}{HTML}{E69F00}
\definecolor{figpaper}{HTML}{F7F9FB}
\usepackage{booktabs}
\usepackage{multirow}
\usepackage{array}
\usepackage{tabularx}

\usepackage{siunitx}
\usepackage{listings}
\usepackage{tikz}
\usetikzlibrary{arrows.meta,positioning,calc,shapes,shapes.geometric,fit,backgrounds}

\usepackage{algorithm}
\usepackage{algpseudocode}

\usepackage[hidelinks]{hyperref}
\usepackage[capitalize]{cleveref}
\makeatletter
\providecommand*{\theHALG@line}{\thealgorithm.\arabic{ALG@line}}
\makeatother
\crefformat{section}{\S#2#1#3}
\Crefformat{section}{Section~#2#1#3}
\crefformat{subsection}{\S#2#1#3}
\Crefformat{subsection}{Section~#2#1#3}

\usepackage{balance}
\usepackage{xspace}
\usepackage{enumitem}

\let\oldtexttt\texttt
\renewcommand{\texttt}[1]{{\small\oldtexttt{#1}}}

\theoremstyle{definition}

\newcommand{\sysname}{GPU-resident Execution\xspace}

\begin{document}

\title{GPU-resident Execution for Timing-predictable Sensor Processing Systems}

\author{
  \IEEEauthorblockN{\makebox[.32\textwidth]{Soham Sinha\textsuperscript{*}}}
  \IEEEauthorblockA{\makebox[.32\textwidth]{\textit{NVIDIA}}\\
                    \makebox[.32\textwidth]{Bengaluru, India}}
  \and
  \IEEEauthorblockN{\makebox[.32\textwidth]{Rahul Choudhury\textsuperscript{*}}}
  \IEEEauthorblockA{\makebox[.32\textwidth]{\textit{NVIDIA}}\\
                    \makebox[.32\textwidth]{Santa Clara, CA, USA}}
  \and
  \IEEEauthorblockN{\makebox[.32\textwidth]{Shekhar Dwivedi}}
  \IEEEauthorblockA{\makebox[.32\textwidth]{\textit{NVIDIA}}\\
                    \makebox[.32\textwidth]{Hyderabad, India}}
  \and[\hfill\break\hfill]
  \IEEEauthorblockN{\makebox[.32\textwidth]{Gregory Lee, Gigon Bae, and}\\
                    \makebox[.32\textwidth]{Anisha Goel}}
  \IEEEauthorblockA{\makebox[.32\textwidth]{\textit{NVIDIA}}\\
                    \makebox[.32\textwidth]{Santa Clara, CA, USA}}
  \and
  \IEEEauthorblockN{\makebox[.32\textwidth]{Elena Agostini}}
  \IEEEauthorblockA{\makebox[.32\textwidth]{\textit{NVIDIA}}\\
                    \makebox[.32\textwidth]{Milan, Italy}\\
                    \makebox[.32\textwidth]{\textsuperscript{*}\{sohams,rahulc,mazizian\}@nvidia.com}}
  \and
  \IEEEauthorblockN{\makebox[.32\textwidth]{Patrick O'Grady, Yosi Levi, and}\\
                    \makebox[.32\textwidth]{Mahdi Azizian\textsuperscript{*}}}
  \IEEEauthorblockA{\makebox[.32\textwidth]{\textit{NVIDIA}}\\
                    \makebox[.32\textwidth]{Santa Clara, CA, USA}}
}

\maketitle
\vspace{-0.8em}
\thispagestyle{plain}
\pagestyle{plain}

\begin{abstract}
Real-time embedded systems increasingly place GPU-accelerated processing of
high-bandwidth sensor
streams inside closed-loop systems, but the CPU (the host) remains the
primary orchestrator of each iteration. The host observes sensor inputs,
submits CUDA work, waits for completion, and commits actuation. This
host-mediated path exposes the closed-loop end-to-end latency to I/O interrupts,
host scheduler decisions, and CUDA API overheads even after
common host-side mitigations such as GPU-memory RDMA or CPU core isolation.
This paper presents
\emph{GPU-resident Execution}, a new execution model for real-time
sensor processing pipelines that keeps I/O and GPU execution
in a long-lived device-resident
loop. Sensor inputs arrive in GPU memory, trigger GPU-side processing,
and reach GPU-visible output endpoints without any host wakeups.
We implement the model in the NVIDIA Holoscan C++ framework and provide reference
components for building GPU-resident applications.
We evaluate our system on NVIDIA GPUs
and SoCs with inference, host-contention, multi-loop,
and streaming camera workloads. On an industry-grade
1080p, 60~Hz real-time camera-to-display application,
\sysname reduces photon-to-display jitter from
$\SI{4.83}{\milli\second}$ to $\SI{0.66}{\milli\second}$
($7.3\times$) relative to a CPU-driven GPU-memory RDMA baseline and
reduces software-processing jitter from $\SI{1.70}{\milli\second}$ to
$\SI{0.01}{\milli\second}$ ($170\times$).
\end{abstract}

\section{Introduction}\label{sec:intro}

Modern sensor data processing systems increasingly ingest high-bandwidth
camera, radar, LiDAR, and IMU streams that demand parallel processing on
accelerators~\cite{chen24endtoendad,roychoudhury23humanoid,
lee22rtbev}.
This makes accelerator execution part of the bounded end-to-end response-time
required by closed-loop systems.
GPUs are now the dominant accelerator, even for such time-critical workloads, in
autonomous
driving~\cite{nvidia-mb-drive,nvidia-hyundai-drive},
drones~\cite{tijtgat17uav,hossain19aerial},
robotics~\cite{bostondynamics-spot-coreio}, medical
devices~\cite{nvidia-holoscan-medical,ahmed25surgical}, and industrial
automation~\cite{nvidia-visual-inspection,liu23surface}. GPUs are ofered in compact SoC form
factors~\cite{nvidia-jetson-modules,nvidia-igx-orin,nvidia-igx-thor,amd-zynq-ultrascale-mpsoc}
with feature-rich software stacks (CUDA/TensorRT, ROCm/MIOpen), camera
drivers, display, and networking support. However, the conventional CPU-driven
GPU usage keeps it
subordinate to the host, complicating timing predictability.

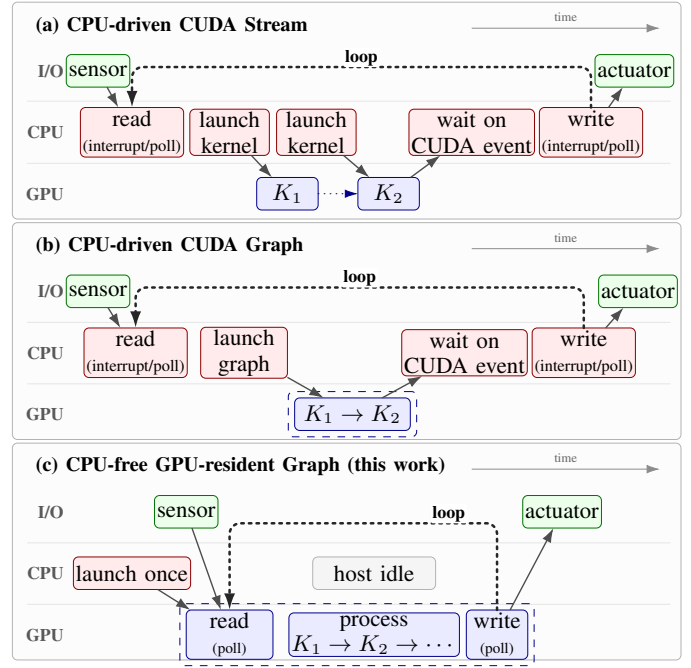
\begin{figure}[t]
  \centering
  \resizebox{\columnwidth}{!}{\begin{tikzpicture}[
    font=\small,
    panel/.style={draw=black!30, rounded corners=2pt, line width=0.35pt,
                  fill=black!1},
    lane/.style={font=\bfseries\scriptsize, anchor=east, text=black!65},
    io/.style={draw=green!45!black, fill=green!8, rounded corners=2pt,
               minimum width=8.5mm, minimum height=4.5mm, inner sep=1pt,
               align=center, line width=0.35pt},
    cpu/.style={draw=red!55!black, fill=red!8, rounded corners=2pt,
                minimum width=8.5mm, minimum height=4.5mm, inner sep=1pt,
                align=center, line width=0.35pt},
    gpu/.style={draw=blue!55!black, fill=blue!8, rounded corners=2pt,
                minimum width=8.5mm, minimum height=4.5mm, inner sep=1pt,
                align=center, line width=0.35pt},
    idle/.style={draw=black!30, fill=black!4, rounded corners=2pt,
                 minimum height=4.5mm, inner sep=1pt, align=center,
                 line width=0.35pt},
    arrow/.style={-{Latex[length=2.0mm,width=1.35mm]}, line width=0.55pt,
                  draw=black!70},
    hostarrow/.style={-{Latex[length=2.0mm,width=1.35mm]}, line width=0.55pt,
                      draw=red!55!black},
    gpuarrow/.style={-{Latex[length=2.0mm,width=1.35mm]}, line width=0.55pt,
                     draw=blue!55!black},
    streamarrow/.style={-{Latex[length=1.8mm,width=1.15mm]}, line width=0.35pt,
                        draw=blue!55!black, dotted, line cap=round},
    looparrow/.style={-{Latex[length=2.3mm,width=1.55mm]}, line width=1.0pt,
                      draw=black!85, dotted, rounded corners=1.5mm,
                      line cap=round},
    looplabel/.style={font=\bfseries\scriptsize, text=black, fill=white,
                      rounded corners=0.8pt, inner sep=0.7pt},
    timeaxis/.style={-{Latex[length=1.8mm,width=1.2mm]}, line width=0.35pt,
                     draw=black!45},
    note/.style={font=\tiny, text=black!72, align=center},
    loopbox/.style={draw=blue!55!black, dashed, rounded corners=2pt,
                    inner sep=2.5pt, line width=0.35pt},
    every node/.append style={outer sep=0pt}
  ]
  \def\panelw{9.25}
  \def\panelh{3}
  \def\xmid{4.62}
  
\node[panel, minimum width=\panelw cm, minimum height=\panelh cm]
    (pa) at (\xmid,0) {};
  \node[font=\bfseries\footnotesize, anchor=west] at (0.20,1.18)
    {(a) CPU-driven CUDA Stream};
  \draw[timeaxis] (6.35,1.14) -- node[above=-0.2mm, font=\tiny, text=black!60]
    {time} (8.95,1.14);
  \draw[black!10] (0.20,0.13) -- (9.10,0.13);
  \draw[black!10] (0.20,-0.73) -- (9.10,-0.73);
  \node[lane] at (0.82,0.55) {I/O};
  \node[lane] at (0.82,-0.30) {CPU};
  \node[lane] at (0.82,-1.15) {GPU};
  \node[io]  (as) at (1.18,0.55) {sensor};
  \node[cpu, minimum width=13mm] (a0) at (1.65,-0.30)
    {read\\{\fontsize{6.5pt}{6.5pt}\selectfont (interrupt/poll)}};
  \node[cpu, minimum width=11mm] (a1) at (3,-0.30) {launch\\kernel};
  \node[cpu, minimum width=11mm] (a3) at (4.2,-0.30) {launch\\kernel};
  \node[gpu] (a2) at (3.8,-1.15) {$K_1$};
  \node[gpu] (a4) at (5.20,-1.15) {$K_2$};
  \node[cpu, minimum width=13mm] (a5) at (6.35,-0.30)
    {wait on\\CUDA event};
  \node[cpu, minimum width=10mm] (a6) at (8,-0.30)
    {write\\{\fontsize{6.5pt}{6.5pt}\selectfont  (interrupt/poll)}};
  \node[io]  (aa) at (8.6,0.55) {actuator};
  \draw[arrow] (as) -- (a0);
  \draw[arrow] (a1) -- (a2);
  \draw[arrow] (a3) -- (a4);
  \draw[streamarrow] (a2) -- (a4);
  \draw[arrow] (a4) -- (a5);
  \draw[arrow] (a6) -- (aa);
  \coordinate (aloopR) at ($(a6.north)+(0,5.6mm)$);
  \coordinate (aloopL) at ($(a0.north)+(0,5.6mm)$);
  \draw[looparrow] (a6.north) -- (aloopR) --
    node[looplabel, above, midway] {loop} (aloopL) -- (a0.north);
  
\node[panel, minimum width=\panelw cm, minimum height=\panelh cm]
    (pb) at (\xmid,-3.05) {};
  \node[font=\bfseries\footnotesize, anchor=west] at (0.20,-1.87)
    {(b) CPU-driven CUDA Graph};
  \draw[timeaxis] (6.35,-1.91) -- node[above=-0.2mm, font=\tiny, text=black!60]
    {time} (8.95,-1.91);
  \draw[black!10] (0.20,-2.93) -- (9.10,-2.93);
  \draw[black!10] (0.20,-3.78) -- (9.10,-3.78);
  \node[lane] at (0.82,-2.50) {I/O};
  \node[lane] at (0.82,-3.35) {CPU};
  \node[lane] at (0.82,-4.20) {GPU};
  \node[io]  (bs) at (1.18,-2.50) {sensor};
  \node[cpu, minimum width=13mm] (b0) at (1.7,-3.35)
    {read\\{\fontsize{6.5pt}{6.5pt}\selectfont  (interrupt/poll)}};
  \node[cpu, minimum width=12mm] (b1) at (3.2,-3.35) {launch\\graph};
  \node[gpu, minimum width=16mm] (b2) at (4.7,-4.20) {$K_1 \rightarrow K_2$};
  \node[cpu, minimum width=13mm] (b3) at (6.25,-3.35)
    {wait on\\CUDA event};
  \node[cpu, minimum width=10mm] (b4) at (7.9,-3.35)
    {write\\{\fontsize{6.5pt}{6.5pt}\selectfont  (interrupt/poll)}};
  \node[io]  (ba) at (8.65,-2.50) {actuator};
  \node[loopbox, fit=(b2)] {};
  \draw[arrow] (bs) -- (b0);
  \draw[arrow] (b1) -- (b2);
  \draw[arrow] (b2) -- (b3);
  \draw[arrow] (b4) -- (ba);
  \coordinate (bloopR) at ($(b4.north)+(0,5.6mm)$);
  \coordinate (bloopL) at ($(b0.north)+(0,5.6mm)$);
  \draw[looparrow] (b4.north) -- (bloopR) --
    node[looplabel, above, midway] {loop} (bloopL) -- (b0.north);

\node[panel, minimum width=\panelw cm, minimum height=\panelh cm]
    (pc) at (\xmid,-6.10) {};
  \node[font=\bfseries\footnotesize, anchor=west] at (0.20,-4.92)
    {(c) CPU-free GPU-resident Graph (this work)};
  \draw[timeaxis] (6.35,-4.96) -- node[above=-0.2mm, font=\tiny, text=black!60]
    {time} (8.95,-4.96);
  \draw[black!10] (0.20,-5.98) -- (9.10,-5.98);
  \draw[black!10] (0.20,-6.83) -- (9.10,-6.83);
  \node[lane] at (0.82,-5.55) {I/O};
  \node[lane] at (0.82,-6.40) {CPU};
  \node[lane] at (0.82,-7.25) {GPU};
  \node[cpu] (c0) at (1.65,-6.40) {launch once};
  \node[io]  (cs) at (2.4,-5.55) {sensor};
  \node[idle, minimum width=17mm] (c1) at (5.00,-6.40) {host idle};
  \node[gpu, minimum width=12mm] (c2) at (3,-7.25)
    {read\\{\fontsize{6pt}{6pt}\selectfont  (poll)}};
  \node[gpu, minimum width=21mm] (c3) at (5,-7.25)
    {process\\$K_1 \rightarrow K_2 \rightarrow \cdots$};
  \node[gpu] (c4) at (6.7,-7.25) {write\\{\fontsize{6pt}{6pt}\selectfont  (poll)}};
  \node[io]  (ca) at (7.6,-5.55) {actuator};
  \node[loopbox, fit=(c2)(c3)(c4)] {};
  \draw[arrow] (c0) -- (c2);
  \draw[arrow] (cs) -- (c2);
  \draw[arrow] (c4) -- (ca);
  \coordinate (cloopR) at ($(c4.north)+(0,12.0mm)$);
  \coordinate (cloopL) at ($(c2.north)+(0,12.0mm)$);
  \draw[looparrow] (c4.north) -- (cloopR) --
    node[looplabel, above, pos=0.18] {loop} (cloopL) -- (c2.north);
  \end{tikzpicture}}
  \caption{Execution models for
  sensing$\to$accelerated processing$\to$actuation loop.}
  \label{fig:execmodels}
  \end{figure}

\textbf{Host-mediated I/O and GPU orchestration remains on the time-critical path.}
In a CPU-driven workflow, the GPU drives accelerated processing but
leaves per-frame control on the host. The CPU observes sensor inputs,
submits GPU work, waits for completion, and commits actuation.
The top two rows of Figure~\ref{fig:execmodels} capture
this host-controlled sequence:
whether the host launches kernels one at a time in a CUDA stream or submits them
as a captured CUDA Graph~\cite{nvidia-cuda-graphs},
per-frame I/O, CUDA operations, and synchronization still route through the CPU.
The sensor-to-output-endpoint latency, therefore, includes host
orchestration, which runs under Linux and is subject to event-delivery,
driver and scheduler timing
variation~\cite{de2022operating,bristot20linux,zhang06interrupt,madden19linuxrt,missimer16io}.

Prior work has attempted to mitigate this host-side cost in Linux.
PREEMPT\_RT patches provide threaded interrupt handlers and makes most kernel
execution preemptible~\cite{linux-rt-theory,reghenzani19preempt,
bristot20linux}; kernel-bypass I/O removes bottlenecks from
the network stack~\cite{voneicken95unet,dpdk}; and core isolation reduces
interference and scheduler preemption~\cite{madden19linuxrt}.
These
techniques improve the environment in which the host runs. They do not
change the control-plane placement: per-frame I/O and CUDA controls
remain in Linux, and end-to-end timing continues
to inherit scheduling delays, kernel overheads, and OS-level uncertainty.

This paper presents \emph{GPU-resident execution}, a new execution model
whose design principle is to avoid the unpredictable CPU components
from the closed-loop system entirely at steady-state
and place them on the GPU
for its lifetime.
We developed it with the observation that
GPU-memory RDMA~\cite{nvidia-gpudirect-rdma,amd-rocm-rdma}
and GPU-driven I/O~\cite{nvidia-gpunetio,amd-rocm-xio} can form a
sensor/actuator substrate, with inputs arriving in GPU-visible memory
and outputs leaving from it.
A device-side handler observes sensor
input,
triggers the data processing pipeline, and writes output signals,
all without a host wakeup (see~\Cref{fig:execmodels}). The CPU remains in
the setup and supervisory plane, so CPU orchestration overheads and
unpredictability do not contribute to the
end-to-end sensor-to-output latency.

This paper is the first to address sensor I/O determinism
and GPU workload execution predictability as one problem, through
a GPU-resident execution model.
We implement a reference C++ framework and examples
for real-time streaming applications.
The framework treats a sensor pipeline as a GPU-native
fixed-execution unit, amenable to GPU timing
analysis and
verification~\cite{betts13wcetgpu,berezovskyi14cuda,jeanmougin25wcetgpu,li12gklee,betts12gpuverify},
rather than a sequence of unbounded CUDA API calls.

We implement \sysname using CUDA Graphs,
GPU-driven I/O and GPU-memory RDMA, and evaluate it on x86,
Orin, and Thor platforms with Ampere, Ada, and Blackwell GPUs.
Across inference, host-contention and scalability
microbenchmarks,
\sysname consistently reduces standard deviation, jitter and
worst-case latency, against a core-isolated CPU-driven baseline using
NVIDIA Holoscan SDK~\cite{holoscan-sdk}. On an industry-grade
1920$\times$1080, 60~Hz IMX274 camera-to-display pipeline, our
framework reduces
photon-to-display jitter by $7.3\times$ over a CPU-driven solution
and software-processing jitter by $170\times$.

The rest of the paper is organized as follows.
\Cref{sec:bg} gives required background for the design.
\Cref{sec:design} presents the system design, API and implementation details of
\sysname. \Cref{sec:ref} describes
the reference operators and IMX274 application, built on our API,
which are evaluated in
\cref{sec:eval}. Finally,~\crefrange{sec:disc}{sec:conc} place the work in scope,
discuss related work, and conclude with future directions.

\section{Background}\label{sec:bg}

ML-driven sensor pipelines on Linux-hosted GPU
platforms~\cite{holoscan-sdk,ros2-foundation,nvidia-deepstream,nvidia-isaac-ros,amd-vvas}
target predictable and strict timing guarantees such as sub-millisecond
end-to-end latency.
This section discusses the challenges that prevent the
CPU-driven execution model from meeting those timing contracts reliably and
then introduces the sensor I/O and CUDA Graph primitives used by
\sysname.

\subsection{Challenges}\label{sec:bg-jitter}

CPU-driven real-time ML pipelines struggle primarily with three systems design challenges.

\paragraph*{(C1)~Unpredictable Host I/O Scheduling}
CPU-driven execution manages sensor I/O events through the Linux I/O stack.
First, an interrupt or polled queue reaches a driver. Then, the kernel notifies
a user thread which is then scheduled before GPU work
is launched, and actuator commands are committed.
This path exposes the time-critical loop to interrupt latency, deferred kernel
work (threaded IRQs, softirqs), polling delay, and
priority mismatch between the I/O event and the real-time userspace
threads~\cite{linux-rt-theory,reghenzani19preempt,
bristot20linux,zhang06interrupt,missimer16io}.

RTOSs can provide stronger I/O timing guarantees through priority-aware
I/O scheduling, budgeted interrupt handling, and domain-isolated OS
designs~\cite{danish11questvcpu,missimer16io,qnx-neutrino,li14questv,sinha21driveos,sinha22thesis},
but supporting the same I/O devices requires rewritten drivers.
In addition, most RTOSs do not
support GPU software stacks (CUDA/TensorRT,
ROCm), camera, advanced display, and advanced networking
drivers.
QNX is the only commercial RTOS with CUDA support~\cite{nvidia-drive-os,
nvidia-drive-qnx-cuda}, and even that is limited to a
safety-certified CUDA subset.
Linux-based mitigations -- threaded interrupt handlers in
PREEMPT\_RT~\cite{reghenzani19preempt,zhang06interrupt},
kernel-bypass I/O~\cite{dpdk,voneicken95unet},
isolated CPU cores for I/O threads -- reduce parts of this cost, but the
critical event path still passes through the Linux kernel and scheduler.

\begin{figure}[t]
\centering
\includegraphics[width=0.9\columnwidth]{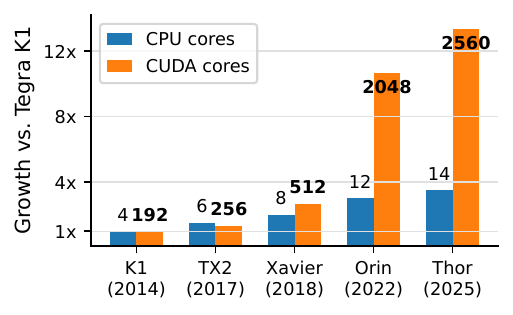}
\caption{NVIDIA SoC CPU and CUDA core count growth~\cite{nvidia-jetson-modules}.}
\label{fig:core_scaling}
\end{figure}

\paragraph*{(C2)~Host Scheduling and Core Isolation Limitations}
On Linux, time-critical I/O and ML applications are managed by
middleware runtimes such as ROS2 and Holoscan, and
share the host with
non-critical, third-party, and background workloads.
Achieving predictable interrupt handling and task release in this
environment requires either an analytically complex or intractable scheduling
configuration, or CPU core isolation that dedicates cores to specific
handlers and threads~\cite{madden19linuxrt}.
Core isolation may improve predictability in some setups, but it
does not scale to modern robotics, humanoid, and industrial workloads
that integrate tens to hundreds of sensors and actuators.
Embedded SoCs provide only a few CPU cores, as shown 
in~\Cref{fig:core_scaling} for NVIDIA platforms,
whereas integrated GPU CUDA core counts already 
outnumber them by over an order of magnitude (PCIe GPUs even more).
This motivates using CUDA cores for real-time compute.

\paragraph*{(C3)~CPU--GPU Coordination Overheads}
In CPU-driven GPU work for ML, each iteration depends on host-mediated
coordination with the GPU: the CPU enters the CUDA runtime/driver,
submits work, and later observes completion through a stream/event
synchronization before the next control decision can execute
(see~\cref{fig:execmodels}). With CUDA Streams, the host submits the
kernel one launch at a time. CUDA Graphs~\cite{nvidia-cuda-graphs}
reduces the number
of launch calls, but the launch and completion path remains part of the
real-time loop. Consequently, CUDA driver queueing and dispatch, the CPU
scheduler, and the kernel notification mechanisms -- all
contribute to timing variation~\cite{olmedo20cuda,
bakita24demystifying,liu24autonomy} and degrade end-to-end predictability.

Together, these challenges motivate \sysname, which keeps the sensor data
path and per-frame data and control flow on the GPU and removes the host from
the critical path. Next, we describe the substrates that make this possible.

\subsection{Sensor I/O}\label{sec:bg-io}

Removing the host from the real-time I/O path requires that sensor data
reach the GPU and actuator commands leave it, directly.
In standard Linux, both directions traverse the kernel or a user
thread. Kernel-bypass frameworks such as DPDK~\cite{dpdk} and
U-Net~\cite{voneicken95unet} map NIC queues directly into user space,
polling receive queues from a dedicated CPU core and reducing
per-packet latency to single-digit microseconds on 100\,Gb/s
connections. GPU-memory
RDMA~\cite{nvidia-gpudirect-rdma,amd-rocm-rdma} extends bypass to the
data path by letting a remote NIC write directly into device memory
over PCIe at microsecond-scale
latency~\cite{nvidia-gpudirect-benchmark,sapio21scaling}. Both
mechanisms remove kernel mediation and CPU copies but leave control on
the CPU.

\subsubsection{Holoscan Sensor Bridge}\label{sec:bg-hsb}

Holoscan Sensor Bridge (HSB)~\cite{holoscan-sensor-bridge,
holoscan-sensor-bridge-github} is an open-source external FPGA
bridge that delivers sensor data to the host.
It packetizes sensor output
as RDMA over Converged Ethernet (RoCEv2) traffic and provides
a unified RDMA/RoCE interface for native sensor links such as MIPI,
SPI, and I2C. ConnectX SmartNICs place frame payloads directly into
pre-allocated device memory via GPU-memory RDMA,
bypassing the host network stack. HSB, therefore, enables a GPU
data-plane for sensor and actuator data, but the CPU
still controls I/O events and frame reception. We preserve
HSB's RDMA data plane and move those decisions into a GPU-resident
data-ready handler (\cref{sec:design-input}) via GPU-driven I/O, placing
both the data path and the steady-state control path on the device.

\subsubsection{GPU-driven I/O}\label{sec:bg-gpunetio}
GPU-driven I/O
frameworks~\cite{nvidia-gpunetio,amd-rocm-xio,agostini17offloading,agostini18gpudirectasync}
go one step further by exposing NIC doorbell registers to device
kernels which can poll for receive completions, post send
requests, and ring doorbells without CPU
involvement~\cite{nvidia-gpunetio-blog,nvidia-gpunetio-rdma,liao24gpucentric}.
Our design combines GPU-memory RDMA with GPU-driven I/O so the sensor
and actuator path no longer needs a host-side I/O thread.

\subsection{CUDA Graphs}\label{sec:bg-graphs}

GPU-visible I/O places sensor data and actuator endpoints on the
device, but uninterrupted processing for new sensor data
must also execute without host mediation. Advanced CUDA Graphs
features provide those GPU-native control mechanisms.

CUDA Graphs~\cite{nvidia-cuda-graphs} allow an application to capture a
sequence of kernel launches, memory copies and other CUDA operations
as a reusable graph. Launching the graph replaces many host-side CUDA calls
with a single call, but in CPU-driven runtimes that call is still issued once
per sensor frame. Indeed, most embedded SDKs using GPUs, including
Isaac ROS~\cite{nvidia-isaac-ros} and
Holoscan SDK~\cite{holoscan-cuda-stream-handling},
use either \emph{CPU-driven} CUDA Streams or repeated
host-launched CUDA Graphs.

Two CUDA primitives can eliminate this per-frame host relaunch.
\emph{(1)}~A CUDA Graph conditional \textsc{while}
node~\cite{nvidia-cuda-graphs-conditional} re-executes a body
subgraph without leaving the device.
\emph{(2)}~Device graph tail-self-launch
repeatedly relaunches the same graph from the
device~\cite{nvidia-device-graph-launch}.
Both mechanisms enable a
frame-processing loop to run entirely on the device.
Unlike any prior work,
\sysname integrates these primitives with
GPU-visible I/O into an execution model for real-time sensor
processing pipelines.

\section{System Design and Implementation}\label{sec:design}

\sysname expresses a sensor processing application as a directed
acyclic graph (DAG) of \emph{GPU-resident operators}. Each operator,
representing a basic functional component of the application,
reads device-visible inputs, runs one or more CUDA kernels, and
writes device-visible outputs. Data-flow edges connect the output
\emph{ports} of one operator to the input \emph{ports} of another.
\sysname encapsulates this operator DAG into
a single GPU-resident CUDA Graph that the host launches once. Next, we describe
the \sysname framework and GPU-resident Graph before discussing the API and implementation details.

\subsection{Framework Architecture and Lifecycle}\label{sec:design-phases}

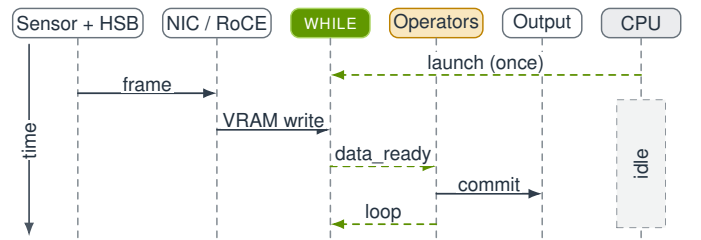
\begin{figure}[H]
  \centering
  \resizebox{\columnwidth}{!}{\begin{tikzpicture}[
    font=\small\sffamily,
    every path/.style={line cap=round, line join=round},
    lifeline/.style={line width=0.72pt, dashed, draw=figink!55},
    msg/.style={-{Latex[length=2.25mm,width=1.65mm]}, line width=0.82pt,
                draw=figink!94},
    msgback/.style={-{Latex[length=2.25mm,width=1.65mm]}, line width=0.82pt,
                    dashed, draw=nvidiagreen!78!black},
    timearrow/.style={-{Latex[length=2.25mm,width=1.65mm]}, line width=0.82pt,
                      draw=figink!82},
    actor/.style={draw=figink!68, rounded corners, minimum width=13mm,
                  minimum height=4.8mm, inner sep=1.8pt, align=center,
                  fill=white, text=figink, line width=0.68pt},
    framework/.style={actor, draw=nvidiagreen!62!white,
                      fill=nvidiagreen!82!black, text=white},
    appwork/.style={actor, draw=figamber!82!black, fill=figamber!25},
    netnode/.style={actor, draw=figamber!82!black, fill=figamber!25},
    hostnode/.style={actor, draw=figink!68, fill=figink!9}
  ]
\node[actor] (sensor) at (0, 0) {Sensor + HSB};
  \node[actor, right=3mm of sensor] (nic) {NIC / RoCE};
  \node[framework, right=3mm of nic] (ctrl) {\textsc{while}};
  \node[appwork, right=3mm of ctrl] (ops) {Operators};
  \node[actor, right=3mm of ops] (drm) {Output};
  \node[hostnode, right=3mm of drm] (cpu) {CPU};
  
\foreach \n in {sensor, nic, ctrl, ops, drm, cpu}
    \draw[lifeline, dashed] (\n.south) -- ++(0,-33.5mm);
  \draw[timearrow] ([xshift=-8mm,yshift=-1mm]sensor.south) --
    node[midway,rotate=90,font=\small\sffamily,text=figink,fill=white,inner sep=0.8pt] {time}
    ([xshift=-8mm,yshift=-32.5mm]sensor.south);
  
\coordinate (t0) at ($(cpu.south)+(0,-6mm)$);
  \draw[msgback] (cpu.south |- t0) -- node[above,font=\small\sffamily,text=figink,fill=white,inner sep=0.7pt]{launch (once)}
              (ctrl.south |- t0);
  
  \coordinate (t1) at ($(sensor.south)+(0,-9mm)$);
  \draw[msg] (sensor.south |- t1) -- node[above,font=\small\sffamily,text=figink,fill=white,inner sep=0.7pt]{frame}
              (nic.south |- t1);
  
  \coordinate (t2) at ($(nic.south)+(0,-15mm)$);
  \draw[msg] (nic.south |- t2) -- node[above,font=\small\sffamily,text=figink,fill=white,inner sep=0.7pt]{VRAM write}
              (ctrl.south |- t2);
  
  \coordinate (t3) at ($(ctrl.south)+(0,-21mm)$);
  \draw[msgback] (ctrl.south |- t3) -- node[above,font=\small\sffamily,text=figink,fill=white,inner sep=0.7pt]{data\_ready}
              (ops.south |- t3);
  
  \coordinate (t4) at ($(ops.south)+(0,-25.5mm)$);
  \draw[msg] (ops.south |- t4) -- node[above,font=\small\sffamily,text=figink,fill=white,inner sep=0.7pt]{commit}
              (drm.south |- t4);
  
  \coordinate (t5) at ($(ctrl.south)+(0,-30.5mm)$);
  \draw[msgback] (ops.south |- t5) -- node[above,font=\small\sffamily,text=figink,fill=white,inner sep=0.7pt]{loop}
              (ctrl.south |- t5);
  
\node[draw=figink!52, dashed, fill=figink!6, text=figink,
        minimum width=8mm, minimum height=21mm,
        anchor=north] at ($(cpu.south)+(0,-10mm)$) {\rotatebox{90}{idle}};
  \end{tikzpicture}}
  \caption{Lifecycle overview (data-flow: solid, control-flow: dashed).}
  \label{fig:seq}
\end{figure}

A GPU-resident application has an initialization phase and a
steady-state phase. The host drives the initialization phase, and the device
runs the steady-state phase. Figure~\ref{fig:seq} shows the lifecycle overview of a
GPU-resident application.
During initialization, the host
configures the sensor and output endpoints, allocates device buffers, constructs
the GPU-resident CUDA Graph from the operator DAG, and launches it once. The
steady-state path is then executed entirely on the GPU.
Each iteration checks for a sensor frame, runs the
operator DAG when input data is ready, and commits the result to the output
endpoint. The host remains outside the time-critical steady-state path.

\begin{algorithm}[H]
  \caption{High-level GPU-resident application lifecycle}
  \label{alg:while}
  {\small
  \renewcommand{\texttt}[1]{{\footnotesize\oldtexttt{#1}}}
  \begin{algorithmic}[1]
  \State \textbf{host:} configure endpoints, device buffers, and operator DAG
  \State \textbf{host:} construct and launch the GPU-resident CUDA Graph once
  \State \textbf{device:} execute the steady-state loop below
  \While{application is active}
    \State check for device-visible sensor input
    \If{input is not ready}
      \State sleep for a configurable interval; \textbf{continue}
    \EndIf
    \State invoke the operator DAG
    \State commit the result to the output endpoint
  \EndWhile
  \end{algorithmic}
  }
\end{algorithm}

\Cref{alg:while} makes the phase boundary explicit. The host performs
the first two steps once, while the GPU repeats the sensor-to-output
loop. To implement this loop, \sysname places a \emph{data-ready
handler} and a \emph{controller} before the application operator DAG.
The handler checks for new input, the controller decides whether to
execute the operator DAG, and an end-marker completes the iteration.
The next subsection explains how these components form the
GPU-resident Graph.

\subsection{GPU-resident Graph}\label{sec:design-graph}

\begin{figure}[H]
  \centering
  \resizebox{\columnwidth}{!}{\begin{tikzpicture}[
    x=1mm,
    y=1mm,
    font=\small\sffamily,
    every node/.style={align=center},
    every path/.style={line cap=round, line join=round},
    kernel/.style={draw=figink!55, rounded corners, minimum height=6mm,
                   minimum width=17mm, inner sep=2pt, align=center,
                   fill=figpaper, text=figink, line width=0.45pt},
    cond/.style  ={draw=figink!45, minimum height=6mm, minimum width=16mm,
                   inner sep=2pt, align=center, fill=figink!4,
                   text=figink, line width=0.45pt},
    whilenode/.style={draw=nvidiagreen!62!white, rounded corners, minimum height=8mm,
                      minimum width=20mm, inner sep=2pt, align=center,
                      fill=nvidiagreen!82!black, text=white, very thick},
    ifnode/.style ={draw=nvidiagreen!62!white, diamond, aspect=2, minimum width=18mm,
                    inner sep=1pt, align=center, fill=nvidiagreen!82!black,
                    text=white, line width=0.55pt},
    term/.style   ={draw=figink!78, rounded corners, minimum height=6mm,
                    minimum width=14mm, inner sep=2pt, align=center,
                    fill=figink!90, text=white, line width=0.45pt},
    op/.style     ={draw=figamber!72!black, rounded corners, minimum height=5.5mm,
                    minimum width=13mm, inner sep=1pt, align=center,
                    fill=figamber!14, text=figink, line width=0.45pt},
    opdag/.style  ={draw=figink!28, rounded corners, fill=nvidiagreen!4,
                    inner xsep=2mm, inner ysep=3mm, line width=0.45pt},
    arrow/.style ={-{Latex[length=2.0mm,width=1.45mm]}, thick, draw=figink!78},
    dotarrow/.style={-{Latex[length=2.0mm,width=1.45mm]}, thick, dashed,
                     draw=nvidiagreen!60!black},
    edgelabel/.style={font=\small\sffamily, text=figink!82, fill=white,
                      inner sep=0.8pt},
    legendbox/.style={draw=figink!26, rounded corners, fill=white, inner sep=2pt,
                      align=left, font=\scriptsize\sffamily, line width=0.45pt}
  ]
\node[whilenode] (whilen) at (0,0)   {\textsc{while}\\node};
  \node[cond]      (rootc)  at (29,0)  {root cond.\\(0/1)};
  \node[term]      (term)   at (54,0)  {terminate};
  
\node[kernel, fill=figamber!14, draw=figamber!72!black] (drh) at (0,-15) {data-ready\\handler};
  \node[kernel, fill=nvidiagreen!82!black, draw=nvidiagreen!62!white, text=white] (ctrl) at (0,-29) {controller\\kernel};
  \node[cond]      (inc)    at (0,-44)  {input-ready\\cond. (0/1)};
  \node[ifnode]    (ifn)    at (29,-44) {\textsc{if}\\node};
  \node[kernel, fill=nvidiagreen!82!black, draw=nvidiagreen!62!white, text=white] (endm) at (29,-65) {end-marker\\kernel};
  
\node[op] (opa) at (55,-36) {Op A};
  \node[op] (opb) at (55,-51) {Op B};
  \node[op] (opn) at (76,-44) {Op C};
  \begin{scope}[on background layer]
    \node[opdag, fit=(opa)(opb)(opn)] (opdagbox) {};
  \end{scope}
  \node[anchor=north east,font=\small,inner sep=1pt]
    at ([xshift=-1mm,yshift=-0.5mm]opdagbox.north east) {Operator DAG};
  
\node[legendbox, minimum width=45mm, minimum height=23mm] (legend) at (60,-16) {};
  \draw[arrow] ([xshift=2mm,yshift=10mm]legend.west) -- ++(6mm,0)
    node[right=1mm,edgelabel,fill=none,inner sep=0pt] {execution flow};
  \draw[dotarrow] ([xshift=2mm,yshift=6mm]legend.west) -- ++(6mm,0)
    node[right=1mm,edgelabel,fill=none,inner sep=0pt] {device flag control};
  \draw[fill=nvidiagreen!82!black, draw=nvidiagreen!62!white, rounded corners=0.7pt]
    ([xshift=2mm,yshift=0.75mm]legend.west) rectangle ++(6mm,2.5mm);
  \node[anchor=west,font=\small] at ([xshift=9mm,yshift=2.25mm]legend.west)
    {framework node/kernel};
  \draw[fill=figamber!14, draw=figamber!72!black, rounded corners=0.7pt]
    ([xshift=2mm,yshift=-4mm]legend.west) rectangle ++(6mm,2.5mm);
  \node[anchor=west,font=\small] at ([xshift=9mm,yshift=-2.75mm]legend.west)
    {user-supplied node/kernel};
  \draw[fill=figink!4, draw=figink!45]
    ([xshift=2mm,yshift=-8mm]legend.west) rectangle ++(6mm,2.5mm);
  \node[anchor=west,font=\small] at ([xshift=9mm,yshift=-6.75mm]legend.west)
    {condition flag};
  
\draw[dotarrow] (rootc.west) -- node[above,edgelabel,font=\small,yshift=1mm]{check} (whilen.east);
\draw[arrow] (whilen.south) -- node[left,edgelabel,font=\small,xshift=-1mm]{true} (drh.north);
\draw[arrow] (whilen.north) -- ++(0,2mm) -|
  node[pos=0.28,below,edgelabel,font=\small,yshift=1.5mm,xshift=-34mm]{false} (term.north);
\draw[arrow] (drh.south) -- (ctrl.north);
\draw[arrow] (ctrl.east) -- ++(18mm,0) -|
     (ifn.north);
  \draw[dotarrow] (ctrl.west) -- ++(-5mm,0) |- (inc.west);
\draw[dotarrow] (ctrl.north east) -- ++(20mm,0) --
    node[pos=0.4,left,edgelabel,font=\small,xshift=-1mm]{tear-down: 0} (rootc.south);
\draw[dotarrow] (inc.east) -- node[above,edgelabel,font=\small,yshift=1mm]{check} (ifn.west);
\draw[arrow] (ifn.east) -- node[above,edgelabel,font=\small,yshift=1mm]{true} (opdagbox.west);
\draw[arrow] (opa.south) -- (opb.north);
  \draw[arrow] (opa.east) -- (opn.north west);
  \draw[arrow] (opb.east) -- (opn.south west);
\draw[arrow] (opn.south) -- ++(0,-18mm) -- (endm.east);
\coordinate (loopx) at ($(endm.west)+(-36mm,0)$);
  \coordinate (falsebend) at ($(ifn.south)+(0,-10mm)$);
  \draw[arrow] (ifn.south) --
    node[pos=0.45,right,edgelabel,font=\small]{false} (falsebend) --
    (loopx |- falsebend);
\draw[arrow] (endm.west) -- (loopx) |-
    node[pos=0.25,right,edgelabel,font=\small,yshift=-22mm]{next\\iteration} (whilen.west);
  \end{tikzpicture}
  }
\caption{Internal structure of the GPU-resident CUDA Graph.}
\label{fig:gr-graph}
\end{figure}
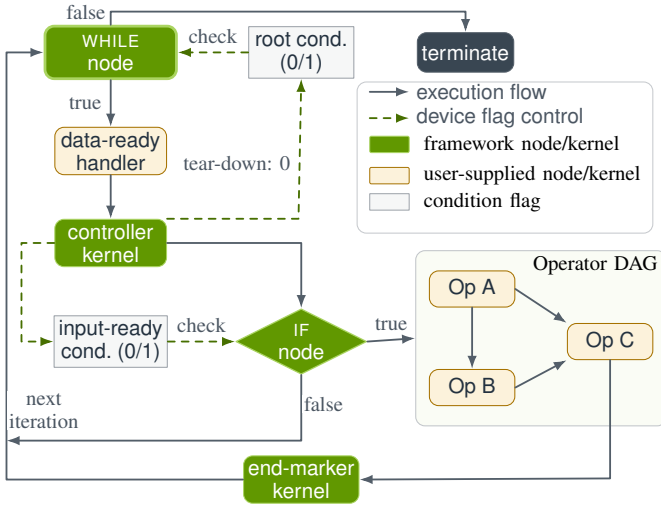

The application creates a GPU-resident CUDA
Graph from its operator DAG.
We provide two mechanisms for keeping this
graph resident on the device: CUDA Graph \textsc{while} conditional
node (WHL)~\cite{nvidia-cuda-graphs-conditional} and device graph tail
launch (DVG)~\cite{nvidia-device-graph-launch}. Their
performance is similar (see~\cref{sec:eval-overhead}), but they differ
in graph-node support and device-resident execution control mechanics.
\Cref{fig:gr-graph} shows the internal structure of our default WHL
mechanism.

The root of the GPU-resident CUDA Graph is a \textsc{while}
conditional node. It keeps the graph active in the device until its condition
becomes false, at which point the graph terminates.
As long as the \textsc{while} condition is true, the application remains active
in device-resident steady-state mode, and the graph executes the
\textsc{while} body. The data-ready handler is the first node
in the body and checks for new input via a user-supplied subgraph of CUDA kernels
(see~\cref{sec:design-drh}).
Next, the controller kernel determines whether to execute the operator
DAG. It sets the predicate of the subsequent \textsc{if} conditional
node from the handler's data-ready flag. This \textsc{if} node gates the
operator DAG. When its predicate is true, the DAG executes before the
end-marker marks the completion of an iteration. CUDA Graph dependencies
enforce this overall execution order without an additional framework scheduler
and timing overheads.

During initialization, \sysname flattens the operator DAG in
topological order. Serialization satisfies all data-flow dependencies
and avoids parallel kernel launches inside the captured graph, which
would introduce timing variability from the on-device work scheduler.
\sysname captures each operator's CUDA work in this order on a
dedicated stream to form the true branch of the \textsc{if} conditional node,
as depicted in~\Cref{fig:gr-graph}.

\subsubsection{GPU-resident Controller}\label{sec:design-controller}
The controller kernel is the device-side decision point in the GPU-resident CUDA Graph.
This single-thread controller translates control flags into the
\textsc{while} and \textsc{if} predicates. It runs after the data-ready
handler in every \textsc{while} iteration. \Cref{alg:controller} shows
pseudocode for the controller kernel.

\begin{algorithm}[H]
  \caption{Controller kernel in WHL CUDA Graph}
  \label{alg:controller}
  {\small
  \algtext*{EndIf}
  \begin{algorithmic}[1]
  \State $d \gets \mathtt{dataReady}$; $t \gets \mathtt{tearDown}$
         \Comment{atomic loads}
  \State $\mathtt{ifPred} \gets (d \ne 0)$
  \If{$d = 0$}
    \State sleep for $\delta\,\si{\micro\second}$
  \EndIf
  \If{$t = 1$}
    \State $\mathtt{whilePred} \gets 0$; $\mathtt{ifPred} \gets 0$
  \EndIf
  \end{algorithmic}
  }
\end{algorithm}

\texttt{tearDown} is a host-mapped control flag through which the host
requests termination of the GPU-resident graph. When set, the controller
clears both predicates, skipping the operator DAG and terminating the
graph. Because the flag is polled once per
iteration, termination takes effect within one loop boundary in the
worst case.

The \texttt{dataReady} flag carries input availability through a
steady-state iteration. The data-ready handler sets it to~1 when input
is available, allowing the controller to release the operator DAG
through the \textsc{if} node. If it remains~0, the controller
clears the \textsc{if} predicate and sleeps for a configurable interval
$\delta\,\si{\micro\second}$.
After the \textsc{if} node's true branch completes, the end-marker resets
\texttt{dataReady} to~0 and sets the \texttt{resultReady} flag.
\texttt{resultReady} is an optional host-mapped completion flag for
debugging or readback and does not participate in steady-state control.
For debugging, the host may map the
\texttt{dataReady} flag and control it.

\subsubsection{Device Graph Tail Launch}\label{sec:design-dvg}
GPU-resident Execution also supports the \emph{device-launchable} CUDA
Graphs~\cite{nvidia-device-graph-launch} mechanism to keep the graph resident on
the device.
A device graph is instantiated with
\texttt{cudaGraphInstantiateFlagDeviceLaunch} and uploaded once to the device.
A running kernel can launch this graph
via \texttt{cudaStreamGraphTailLaunch}, without host involvement.
\sysname builds two such graphs.
\emph{Graph~M} (control) holds the data-ready handler followed by a single-thread
\emph{dispatcher} kernel that replaces the WHL controller (\Cref{alg:controller}).
\emph{Graph~W} (workload) sequentially stitches the captured operator DAG, the
end-marker kernel, and a single-thread \emph{tail-relaunch} kernel.
The host fires Graph~M once via \texttt{cudaGraphLaunch}; every
subsequent iteration boundary is a device-side tail launch, iterating
as in the \textsc{while} design.

The dispatcher operates on the same flags as the WHL controller:
\texttt{tearDown} stops the execution, \texttt{dataReady}~=~1
tail-launches Graph~W, and \texttt{dataReady}~=~0 sleeps for a
configurable interval then tail-launches Graph~M. At the end of
Graph~W, a tail-relaunch kernel unconditionally tail-launches
Graph~M, closing the loop. All flag contracts and the data-ready
handler are reused unchanged.

Because device-launchable graphs may
not contain conditional nodes, this mechanism cannot support conditional
subgraphs that CUDA-based libraries increasingly
adopt~\cite{pytorch-cuda-graph-cond}. Therefore, it is not
default in \sysname, but it provides a path to
low-overhead and flexible on-device dynamic scheduling in the
future. Both mechanisms expose the same application-facing API
described next.

\subsection{Application, Operator, and Handler API}\label{sec:design-api}

The GPU-resident graph above is constructed from three user-facing
abstractions. \sysname exposes them through a C++/CUDA API integrated
into the NVIDIA Holoscan SDK~\cite{holoscan-sdk,holoscan-gpu-resident}.
An \texttt{Application} defines the operator DAG and controls its
lifecycle. An operator is Holoscan's basic functional unit.
\sysname represents it as a
\texttt{GPUResidentOperator}, which implements a processing or I/O
stage. A \texttt{GPUResidentHandler} defines an operator subgraph that
detects new device-visible input. The following interfaces describe
these roles before \Cref{lst:api} combines them in an example.

\subsubsection{Application Interface}
The application's \texttt{compose()} method instantiates operators with
\texttt{make\_operator()} and connects them with \texttt{add\_flow()}.
A handler's \texttt{compose()} method uses the same API to form a DAG
of input-detection operators. The application's
\texttt{gpu\_resident()} accessor exposes
\texttt{register\_data\_ready\_handler()}, which attaches this handler
to the application. Composition and handler registration occur during
host-side initialization.

After composition, \texttt{run\_async()} constructs and launches the
GPU-resident Graph once, beginning device-resident steady-state streaming. While
it runs, the host can query its state with \texttt{is\_launched()} and
request shutdown with \texttt{tear\_down()}.

\subsubsection{Operator Interface}
A \texttt{GPUResidentOperator} performs only CUDA operations in \sysname.
These CUDA operations are captured from the host once, and are repeated as part of the
GPU-resident Graph, as explained in~\cref{sec:design-graph}.

Following the Holoscan convention, \texttt{setup(OperatorSpec\&)}
declares an operator's configuration. \sysname extends
\texttt{OperatorSpec} with
\texttt{device\_input("name", n\_bytes)} and
\texttt{device\_output("name", n\_bytes)} methods to declare named device-side
ports and their storage requirements.
Each port uses one of two device-memory modes. With
\textit{internally-managed} memory, the executor allocates one shared
buffer per edge, maps it to both endpoints, and manages its lifetime.
With \textit{externally-managed} memory, an endpoint supplies a
preallocated device pointer that is mapped to both endpoints. If both
endpoints supply pointers, the source pointer is used.

During graph construction, operators are visited in topological
order. It calls \texttt{setup()} to declare ports and bind parameters,
then resolves each edge to a device buffer. It calls \texttt{compute()}
once in the same order. This method obtains port pointers through
\texttt{device\_memory("name")} and issues CUDA operations on
\texttt{cuda\_stream()}. These CUDA operations are recorded into
a CUDA Graph as a subgraph which is inserted in the \textsc{if} node's true branch
(see~\cref{fig:gr-graph}).

\subsubsection{Data-ready Handler Interface}\label{sec:design-drh}

A data-ready handler performs device-side input detection at the start
of every GPU-resident iteration. Its
\texttt{GPUResidentOperator} subgraph may read sensor metadata, poll a
NIC completion queue, or otherwise inspect device-visible input state.
When input is available, a kernel signals it via the \texttt{dataReady} flag.

During graph construction, handler operators issue their CUDA operations on
\texttt{data\_ready\_handler\_cuda\_stream()}. The
handler subgraph is recorded and inserted as the first node in the
\textsc{while} body (see~\cref{fig:gr-graph}). 
Reference handlers are described
in~\cref{sec:ref-handlers}.

\begingroup
\begin{lstlisting}[language=C++,basicstyle=\ttfamily\scriptsize,
  caption={GPU-resident application example},
  label=lst:api]
// processing operator
class MyKernelOp : public GPUResidentOperator {
  void setup(OperatorSpec& s) override {
    s.device_input ("in",  N*sizeof(float));
    s.device_output("out", N*sizeof(float));
  }
  void compute() override {
    auto *in  = (float*)device_memory("in");
    auto *out = (float*)device_memory("out");
    my_kernel<<<G,B,0,cuda_stream()>>>(in,out);
  }
};
class GpuResidentApp : public Application {
  void compose() override {
    auto sense = make_operator<SensorOp>("src");
    auto k1  = make_operator<MyKernelOp>("k1");
    auto output = make_operator<OutputOp>("sink");
    add_flow(sense, k1);  add_flow(k1, output);
  }
};
// data-ready handler
class MyDRH : public GPUResidentHandler {
  void compose() override {
    add_operator(make_operator<HsbPollOp>("drh"));
  }
};
int main() {
  auto app = make_application<GpuResidentApp>();
  auto drh = make_handler<MyDRH>();
  app->gpu_resident().register_data_ready_handler(drh);
  auto fut = app->run_async();
  // GPU-resident Graph launched, CPU free for other work.
  app->gpu_resident().tear_down();  fut.wait();
}
\end{lstlisting}
\endgroup
\vspace{-4pt}
\subsection{Static SM Affinity for GPU-resident Graphs}\label{sec:design-concurrency}

Each GPU-resident application runs inside a CUDA context assigned a
full GPU or a static CUDA MPS partition~\cite{nvidia-mps-static-sm}.
The partition fixes the streaming multiprocessor (SM)
budget available to the graph for its lifetime,
and all
data-ready, controller, operator, and egress kernels execute in that
context on each iteration. Predictable GPU-resident execution also
requires locked GPU clocks during deployment (e.g.,
\texttt{nvidia-smi --lock-gpu-clocks}~\cite{nvidia-smi}), so the
steady-state loop runs on a fixed resource at a stable frequency. This
forms a key practical basis for timing predictability.
We do not currently consider shared L2 cache, memory bandwidth,
and other device-side interference, whose control would require stronger resource
partitioning~\cite{jain19fractional,zhang25sgdrc}, akin to Intel RDT~\cite{intel-rdt}.

\vspace{-4pt}
\subsection{Timing Predictability by Design}\label{sec:design-timing}

The timing consequence of the design is straightforward: once the graph has been
launched, every steady-state decision that releases work for a sensor
sample is made inside the device-resident graph. We use CUDA Graph
dependencies to establish a single device-side chain from ingress to
egress. When the
data-ready handler marks input ready, the
controller updates the \textsc{if} predicate before that \textsc{if} node
is reached in the same loop iteration. The processing body is therefore
released without a host wakeup, a host-side completion event, or a
per-frame \texttt{cudaGraphLaunch} call by the host.

This gives a design-level timing contract rather than a new GPU
scheduler. The deployed loop must have a fixed graph topology,
preallocated device buffers, locked clocks, and fixed compute resources
assigned to the CUDA context. Under that deployment model, the
steady-state response path has only two kinds of costs: device-side execution
inside the graph and the fixed graph transitions between stages.

If a sample arrives just after the
handler has checked for input, the critical-instant wait is the time
until the next handler check: one controller pass plus the larger of the
configurable data-not-ready sleep interval and the
operator kernels' worst-case execution time.
After the next handler observes the sample, the body release and commit
remain inside that same graph iteration.

This timing model makes overload behavior explicit. If a sensor or
actuator runs faster than the body can sustainably process, \sysname does
not hide the mismatch behind an implicit queueing policy. The
application defines the over-rate behavior at the data-ready handler and
egress boundary: it may process the newest available sample and drop
older ones, suppress duplicate processing when the loop polls faster
than the sensor produces data, or repeat the latest output when an
actuator commits faster than new processed results are available. The
timing model then applies to the samples that this application policy
elects to process.

Thus the main timing claim is not that GPU work is inherently
deterministic, but that the per-frame host scheduler, interrupt, and
CUDA-launch terms are absent from the steady-state response path; the
remaining terms are device-side graph and operator costs measured in the
evaluation.

\section{Reference Operators, Handlers and Applications}\label{sec:ref}

This section presents reference implementations of the framework's ingress,
processing, and egress components and then combines them in a real-time
camera-to-display application.

\vspace{-4pt}
\subsection{GPU-resident Operator Implementations}\label{sec:ref-builtins}\label{sec:design-builtins}

These reference operators show that the API in
\cref{sec:design-api} supports custom CUDA kernels and library-backed
image-processing and inference stages. They
configure image dimensions, tensor layouts, auxiliary device buffers,
library contexts, and kernel launch parameters before their CUDA
operations are captured into a CUDA Graph.

\textbf{Bayer Demosaic.}
Bayer is the raw color-filter-array layout produced by the camera, and
demosaicing reconstructs a full-color image from it. This operator
wraps NVIDIA Performance Primitives (NPP) demosaicing
functions~\cite{nvidia-npp}. It takes image size, pixel type, Bayer layout, and
alpha as setup parameters, and
sets the final \texttt{in} and \texttt{out}
port sizes during initialization. Graph construction then allocates
the corresponding device-side input and output buffers.
For a 1920$\times$1080 raw image with 16-bit
Bayer input and four RGBA output channels, the ports are 4.1\,MB and
16.6\,MB. During capture, the operator binds the NPP stream context to
\texttt{cuda\_stream()} and calls \texttt{nppiCFAToRGB\_*} or
\texttt{nppiCFAToRGBA\_*} kernels. The NPP demosaic operation is captured as a
CUDA Graph node, so the host does not resubmit it for each frame.

\textbf{Format Conversion.}
The format conversion operator supports fixed device-only packed and
planar layouts. It fixes input and output pixel formats, image
dimensions, optional resize parameters, plane strides, and plane
offsets before launch, and allocates any resize or channel-reorder
scratch buffers during initialization. During capture it calls the
corresponding NPP conversion on \texttt{cuda\_stream()}, so color
conversion and optional resize remain inside the graph body.

\textbf{TensorRT Inference.}
The inference operator places model inference in the same captured body
as hand-written kernels. It derives input and output port sizes
from the model configuration, constructs a TensorRT-backed inference
context, and connects the context's input and output buffers to
executor-owned device-port addresses. Once configured, TensorRT exposes
the model as a reusable CUDA Graph, which the operator's
\texttt{compute()} launches on
\texttt{cuda\_stream()} during capture.

\textbf{Camera Image Operators.}
Together, these operators reconstruct and correct a raw camera frame
before demosaicing.
\textit{1) Camera-frame reconstruction.}
This operator reassembles a MIPI Camera Serial Interface 2 (CSI-2)
packet stream into a dense 16-bit Bayer frame for RAW8, RAW10, and
RAW12 sensor formats. The input
port is sized to the CSI byte length and the output port to one 16-bit
Bayer value per pixel.
\textit{2) Raw-image signal processing.}
This operator applies sensor-level corrections in-place, by mapping both
\texttt{in} and \texttt{out} to a single internal device buffer.
Initialization derives Bayer-pattern channel offsets, computes the
optical-black shift from the pixel format, and allocates a
three-channel 256-bin histogram with a white-balance gain vector that
stores the per-channel scale factors. During
capture, the operator runs four kernels on the same stream:
black-level correction, histogram accumulation, white-balance gain
computation, and the final per-pixel correction pass.

\vspace{-4pt}
\subsection{Holoscan Sensor Bridge (HSB) Data-ready Handlers}\label{sec:ref-handlers}\label{sec:design-input}

HSB is an external sensor-side FPGA bridge that packetizes native
sensor data into RoCE traffic for a ConnectX NIC~\cite{holoscan-sensor-bridge}. 
A data-ready handler is required to integrate HSB sensor stream into a
GPU-resident application. This handler
detects new frames on the GPU and signals input availability to the controller kernel.
Implementing such a handler requires the underlying I/O interface to
expose data buffers
and completion state and/or doorbell registers to the GPU.
We implement two HSB-facing handlers
using GPUDirect RDMA and GPU-driven networking with DOCA GPUNetIO~\cite{nvidia-gpunetio}.
Other sensor interfaces can use the same abstraction by exposing this
contract directly or through a bridge like HSB.

\textbf{HSB metadata handler.}
This first handler detects frames from GPU-visible metadata while a CPU
thread continues to manage the NIC receive completion queue.
HSB connects to a MIPI CSI-2 camera, such as the
IMX274~\cite{leopard-imx274-mipi-cs}. It receives the camera stream
and sends each frame as chunked RoCEv2 RDMA Write requests. The
ConnectX NIC places the payload into a two-slot
(or larger, sensor rate-dependent)
GPU memory buffer via DMA. After each
frame, the FPGA sends a metadata packet (frame counter, status,
timestamps) as an RDMA Write with Immediate~\cite{ibta-spec}. This
generates a completion event~\cite{nvidia-rdma-aware-programming} on the ConnectX
NIC~\cite{nvidia-connectx7}. In RDMA terminology, a Completion Queue
(CQ) reports finished work requests, while a Queue Pair (QP) is the
send/receive endpoint that posts them. In the canonical CPU-driven path, a CPU
thread polls the receive CQ, copies metadata to host memory, selects
the frame on the CPU, and releases
downstream processing.

The GPU-resident handler, instead, uses a CUDA
kernel that polls two metadata records directly on the GPU and selects
the slot with the newer frame counter. On a new frame, the handler sets
the \texttt{dataReady}
flag and triggers the rest of the pipeline. This removes the
device-to-host metadata copy and CPU frame-selection decision from
steady state. A lightweight CPU thread still manages the receive CQ,
but frame detection and application execution proceed on the GPU.
The next handler removes the remaining host role.

\textbf{HSB DOCA GPUNetIO handler.}
This second handler removes the remaining per-frame CPU I/O role. The
GPU polls receive completions and reposts receive descriptors through
the NVIDIA DOCA GPUNetIO API~\cite{nvidia-gpunetio}.
During initialization, the host uses DOCA GPUNetIO to allocate the CQ
rings, QP work queues, doorbell record, and receive-frame ring in GPU
memory. The ConnectX driver exposes controls that bind these GPU memory
allocations to the NIC. The host then exports the QP as a device-visible handle and
connects it to the HSB FPGA peer. A setup CUDA kernel pre-posts receive Work Queue
Elements (WQEs) identifying the GPU buffers available to the NIC. Thus, the
NIC driver remains on the host, but the steady-state control is moved
to the GPU.

\sysname captures the DOCA GPUNetIO kernel launch in the handler's
\texttt{compute()} method
on the data-ready handler CUDA stream.
In the streaming phase, the kernel
performs a non-blocking poll of the receive CQ using the DOCA
GPUNetIO
API~\cite{nvidia-gpunetio}. On a miss, it writes
\emph{data not ready}. On a hit, it reads the immediate-data field of
the Completion Queue Entry (CQE) to identify the completed receive-buffer
slot, then copies that frame to a device buffer for processing.
It then advances the CQ consumer position, posts the next receive WQE, and signals
\emph{data ready}.

\vspace{-4pt}
\subsection{Display Output}\label{sec:ref-egress}\label{sec:design-output}
An output operator closes the end-to-end loop by committing the final
device-resident result to an externally visible action. Our display
operator demonstrates this egress pattern via rendering an image stream to a monitor, analogous to
a network transmit or GPIO write.
During initialization, it fixes the display mode (resolution,
frame rate, etc.), creates the
display swapchain, and obtains device pointers to the scanout buffers
using the Linux DRM API~\cite{linux-drm}. It does not create
a Vulkan, OpenGL, or other graphics context, avoiding a costly
graphics/compute context switch~\cite{sinha24deterministic}
on every frame. During steady-state, 
the captured graph issues a gamma-corrected display kernel. Then, the
rendering to a monitor happens in one of the two modes.
Front-buffer rendering mode writes one buffer while the display engine
continuously scans it out. Continuous-flip mode alternates two
buffers, scanning out one while rendering into the other.

Other actuators may follow the same structure when they expose a
GPU-visible doorbell. Initialization fixes the queue, register, or
buffer mapping. The captured \texttt{compute()} body writes the payload
and rings the doorbell after processing completes. Supporting an
egress protocol requires vendor-specific integration and is part of ongoing work.

\subsection{IMX274 Camera-to-Display Application}\label{sec:ref-imx274}
\vspace{-2pt}
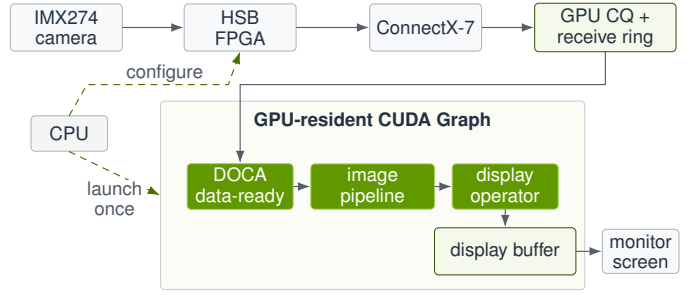
\begin{figure}[!t]
\centering
\resizebox{\columnwidth}{!}{\begin{tikzpicture}[
  font=\small\sffamily,
  every path/.style={line cap=round, line join=round},
  every node/.style={align=center},
  block/.style={draw=figink!38, rounded corners=2pt, minimum height=8mm,
                inner sep=2.4pt, line width=0.45pt, fill=figpaper,
                text=figink},
  hw/.style={block, fill=figpaper, minimum width=19mm},
  net/.style={block, draw=figamber!72!black, fill=figamber!14,
              minimum width=20mm},
  mem/.style={block, draw=nvidiagreen!45!black, fill=nvidiagreen!8,
              minimum width=24mm},
  graph/.style={draw=figink!26, rounded corners=2pt, fill=nvidiagreen!4,
                minimum width=72mm, minimum height=32mm,
                inner sep=2.5pt, line width=0.45pt},
  grstage/.style={draw=nvidiagreen!62!white, fill=nvidiagreen!82!black, rounded corners=2pt,
                  minimum width=18mm, minimum height=7mm,
                  inner sep=1.8pt, line width=0.45pt, text=white},
  cpu/.style={block, fill=figink!5, minimum width=25mm},
  arrow/.style={-{Latex[length=2.1mm,width=1.5mm]}, line width=0.55pt,
                draw=figink!78},
  dotarrow/.style={-{Latex[length=2.1mm,width=1.5mm]}, line width=0.55pt,
                   dashed, draw=nvidiagreen!60!black}
]
\node[hw]  (sensor) at (0,1.75) {IMX274\\camera};
\node[hw]  (hsb)    at (2.95,1.75) {HSB\\FPGA};
\node[hw]  (nic)    at (6.10,1.75) {ConnectX-7};
\node[mem] (ring)   at (9.15,1.75) {GPU CQ +\\receive ring};

\node[graph] (graphbody) at (5.20,-1.10) {};
\node[font=\bfseries\small\sffamily, text=figink, anchor=north] at ($(graphbody.north)+(0,-0.8mm)$)
  {GPU-resident CUDA Graph};
\node[grstage] (drh) at (2.95,-0.95) {DOCA\\data-ready};
\node[grstage, minimum width=21mm] (pipe) at (5.20,-0.95) {image\\pipeline};
\node[grstage] (buf) at (7.45,-0.95) {display\\operator};
\node[mem] (dispbuf) at (7.45,-2.05) {display buffer};

\node[cpu, minimum width=13mm, minimum height=6mm] (host) at (0.05,-0.05) {CPU};
\node[hw, minimum width=13mm, minimum height=6mm] (monitor) at (9.75,-2.05) {monitor\\screen};

\draw[arrow] (sensor) -- (hsb);
\draw[arrow] (hsb) -- (nic);
\draw[arrow] (nic) -- (ring);
\draw[arrow] (ring.south) -- ++(0,-0.55) -| (drh.north);
\draw[arrow] (drh) -- (pipe);
\draw[arrow] (pipe) -- (buf);
\draw[arrow] (buf.south) -- (dispbuf.north);
\draw[arrow] (dispbuf.east) -- (monitor.west);
\draw[dotarrow] (host.north) -- (0.35,0.78) -- node[above, pos=0.55, fill=white, text=figink!82, inner sep=0.7pt]
  {configure} (2.75,0.78) -- (hsb.south);
\draw[dotarrow] (host.south) -- node[below, yshift=-1mm, fill=white, text=figink!82, inner sep=0.7pt]
  {launch\\once} (graphbody.west);
\end{tikzpicture}}
\caption{GPU-resident IMX274 camera-to-display application.}
\label{fig:imx274-app}
\end{figure}

This application combines the ingress handlers, processing operators,
and display egress into a complete externally observable
sensor-to-output loop. \Cref{fig:imx274-app} shows its hardware and
GPU-resident graph structure.
The IMX274 camera~\cite{leopard-imx274-mipi-cs} is the sensor
endpoint in the pipeline, and a monitor is the output endpoint.
The sensor supports 1080p and 4K resolutions at 60~Hz in
10- or 12-bit Bayer RAW format.
A HSB FPGA packetizes each frame as RoCEv2 traffic. A ConnectX-7 (CX7)
NIC on the IGX~Orin~+~RTX~6000~Ada platform delivers the payload
directly into GPU memory.
The existing \emph{CPU-driven} application uses GPUDirect RDMA for
host-bypassed data movement, but the host still handles
HSB-CX7 I/O events, launches CUDA work, and orchestrates each frame.
We build two GPU-resident configurations using the HSB metadata and
DOCA GPUNetIO handlers described above.
For each camera frame, the graph reconstructs and corrects the raw
image, demosaics and converts it, then commits it to the
display, using the previously described operators~\cite{holohub-imx274-gpu-resident}.

\section{Evaluation}\label{sec:eval}

We evaluate \sysname in two steps. First, a set of microbenchmarks
isolates where predictability is gained: one inference pipeline, the
same loop under host contention, many concurrent loops, and the fixed
framework cost of keeping the loop resident on the GPU. Second, an
industry-grade IMX274 camera-to-display case study tests whether
those gains carry through an end-to-end pipeline.

\vspace{-4pt}
\subsection{Setup}\label{sec:eval-setup}

We evaluated \sysname on several platforms and report representative
configurations. All microbenchmarks in
\cref{sec:eval-micro} run on a 24-core/48-hardware-thread 
AMD~7965WX x86\_64 workstation
(Linux kernel 6.17 PREEMPT\_RT) with an RTX~A6000
Ampere (84 SMs, 48 GB VRAM) GPU.
We repeat an experiment on IGX Thor with an integrated GPU
(iGPU: 20 SMs, 128 GB RAM, 14-core Armv9-A CPU, Linux 6.8 PREEMPT\_RT).
The real-time IMX274 streaming camera
application (explained in~\cref{sec:ref-imx274}) case study in
\cref{sec:eval-imx274} runs on IGX~Orin with a 12-core Armv8 CPU
(Linux 5.15 with PREEMPT\_RT), plus an
RTX~6000~Ada (142 SMs, 48 GB VRAM) GPU.

\paragraph{Metrics}
We focus on latency tails and variation as our timing
predictability metrics. A key metric is
absolute finishing
jitter~\cite{buttazzo11hard,cervin04jittermargin}:
for job $k$ with release $r_k$ and
finish $f_k$, response time $R_k=f_k-r_k$ and
jitter $J=\max_k R_k-\min_k R_k$. Microbenchmarks treat each loop iteration as
one job with zero release jitter ($r_k$ at iteration start, $f_k$ at
completion), so $J$ captures end-to-end variation.

\subsection{Microbenchmarks}\label{sec:eval-micro}

The microbenchmarks demonstrate that \sysname improves timing predictability
over CPU-driven baselines against
different sources of timing variation. From a quiet system to
host contention and multi-loop scaling, CPU-driven execution and its
mitigations remain sensitive to host-side effects, while GPU-resident stays
consistently robust. At the end, we measure the fixed
framework overhead.

For CPU-driven baselines, we use Holoscan SDK~\cite{holoscan-sdk}
with CUDA Graphs
rather than individual CUDA
stream launches because Graphs are already
the optimized, lower-overhead host-driven option.
To avoid any host-side scheduling noise, the host process is pinned
to an isolated CPU core configured with
\texttt{isolcpus} (exclude scheduler tasks), \texttt{nohz\_full}
(suppress periodic ticks), and \texttt{rcu\_nocbs} (offload RCU
callbacks), with IRQs offloaded to other cores.

We use box-and-whisker plots to show latency distributions.
Each box spans the middle 50\% of observations
(25th--75th percentiles), the center line marks the median, the diamond
marks the mean, and the whiskers extend to the observed minimum and maximum.

\subsubsection{Single-stage Inference Pipeline}\label{sec:eval-inference}

The first experiment runs a single TensorRT inference operator as the processing
workload in a closed loop, resembling a small ML perception pipeline.
Five TensorRT image-classification CNNs
(ShuffleNetV2-x0.5~\cite{ma18shufflenetv2},
MobileNetV2~\cite{sandler18mobilenetv2},
EfficientNet-B0~\cite{tan19efficientnet},
ResNet-50~\cite{he16resnet}, and
VGG-16~\cite{simonyan15vgg}) are tested.
The two execution models launch the same inference CUDA Graph:
the \textsf{Isolated CPU-driven CUDA Graph} configuration
(\textsf{CPU-CG} hereafter) captures a TensorRT engine once and
host-launches it on every iteration per new input, while the \textsf{GPU-resident}
execution
runs the same model with
\texttt{GPUResidentInferenceOp}, with a data-ready handler to iterate over new
input on the device. Both configurations preload 100
ImageNet images in device memory and round-robin over them during
measurement to simulate a camera stream delivered over GPU-memory
RDMA. For every measurement, we record 50,000 iterations after a 5,000 iteration
warmup.

\begin{figure*}[!t]
\centering
\begin{minipage}[t]{0.63\textwidth}
  \vspace{0pt}
  \centering
  \includegraphics[width=\linewidth]{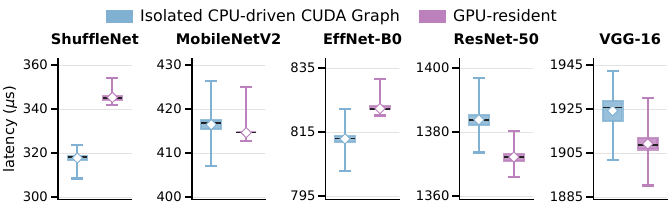}
  \par
  \caption{Single-stage inference latency.}
  \label{fig:exp-e-inference}
\end{minipage}\hfill
\begin{minipage}[t]{0.35\textwidth}
  \vspace{0pt}
  \centering
  \includegraphics[width=\linewidth]{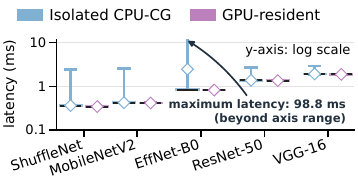}
  \par
  \caption{Single-stage inference under host-side background workloads on RTX A6000.}
  \label{fig:contention-inference}
\end{minipage}
\end{figure*}

\Cref{fig:exp-e-inference} shows the per-iteration latency
distributions.
The experiment is intentionally quiet, and the two modes have comparable
tail latencies. The benefit of \textsf{GPU-resident} appears mainly in lower
standard deviation and a shorter absolute jitter (minimum-to-maximum span).
For MobileNetV2, EfficientNet-B0, and ResNet-50, \sysname reduces
standard deviation and absolute finishing jitter by $1.4\times$--$2.4\times$.
Launching these models as CUDA Graphs per iteration adds variation to the 
\textsf{CPU-CG} timing, even though the launcher is running on an isolated core.
The gains are smaller for ShuffleNetV2 and VGG-16,
where model-level behavior dominates the latency
cost~\cite{kosaian21arithmetic}, and the variance is similar between the two
modes.
\sysname remains subject to
device-side microarchitectural effects such as runtime cache
residency and memory-system contention, but CPU-driven baselines add
host-side overheads on top. 
The next experiments explore more of these host-side
effects.

\subsubsection{Host-side Contention}\label{sec:eval-contention}

In a realistic deployment, the CPU not only runs the real-time loop but
also telemetry logging, health monitoring, non-critical services, and
third-party processes.
We emulate this co-running activity on non-isolated
cores with buffered file I/O ($20\times$ disk I/O) and
memory-intensive workloads (stress-ng~\cite{stress-ng} malloc, vm).
The \textsf{CPU-CG} launcher remains pinned to a fully isolated core,
so the experiment tests interference through shared OS kernel, driver,
and memory subsystem paths, rather than from CPU scheduling.

\textbf{Inference.}
We test against the same five inference models from the previous experiment.
We also use the same iteration regime.
\Cref{fig:contention-inference} shows the latency distributions
for the two configurations in this experiment.
The boxes capture each model's typical latency range, which remains similar
between the two configurations. However, the \textsf{CPU-CG} upper
whiskers extend to the observed maxima and reveal long-tail response-time
outliers. Across four models, \textsf{CPU-CG} jitter expands to
	\SIrange{1.05}{2.15}{\milli\second},
up to an order of magnitude larger than \textsf{GPU-resident},
while EfficientNet-B0 reaches $\SI{97.99}{\milli\second}$,
catching a rare host-side stall, although its median stays low at
$\SI{832}{\micro\second}$.
\textsf{GPU-resident}, in contrast, remains close to the quiet experiment
	with \SIrange{9}{86}{\micro\second} jitter across the five models.
This \textsf{CPU-CG} tail behavior was reproducible across runs, not a
one-off measurement artifact. More broadly, the larger jitter and
corresponding worst-case latencies indicate that a process,
even on an isolated core, is
not immune to host-side contention.
Such latency anomalies did not appear in \sysname  across multiple runs.

Core isolation avoids scheduler preemption of the \textsf{CPU-CG}
process.
However, every iteration still issues CUDA host API calls
which have no timing guarantees in CUDA's user- and kernel-space
drivers.
In addition,
background memory and I/O pressure can affect these calls even on isolated cores,
for example via TLB shootdowns induced by stress-ng, and delayed driver or IRQ
activity. Because of the timing-predictable design of \sysname, a device-resident Graph is immune to such
uncontrollable timing
effects on the host.

\textbf{Emulated I/O.}
Next, we design a lightweight closed-loop emulated I/O
experiment and put it under the same background workloads.
To emulate a sensor/actuator pair, two
SCHED\_FIFO (99 priority) CPU threads with small, locked memory footprints
are pinned to isolated cores: the sensor thread releases a 1~kHz frame token and
timestamps it, while the actuator thread busy-polls the completion
and timestamps the response.
The processing component between the sensor and actuator can be CPU-driven
or GPU-resident.
The latency is the difference between the actuator's and sensor's timestamps.

For this experiment, the CPU-driven configuration is designed as a
kernel-bypass GPU-memory-RDMA-style deployment: the input token is visible in a
page-locked (pinned) buffer, but host-side control still polls it,
launches a CUDA Graph, waits for completion, and emits the output token.
The GPU-resident configuration uses the same emulated endpoints,
but the data-ready handler polls the device-mapped and pinned input token from the
GPU, and the final graph node writes the output token back to host-visible
memory, modeling GPU-memory RDMA plus GPU-driven I/O.
Unlike the previous experiments, latency here is observed by external threads.
For processing workload, we use six 512$\times$512 fp32 matrix multiplication kernels
that run under a millisecond on RTX A6000. For this lightweight workload,
each configuration runs for 1,000,000 loop iterations (first 1,000 dropped
as warmup).

\begin{figure}[!b]
\centering
\includegraphics[width=\columnwidth]{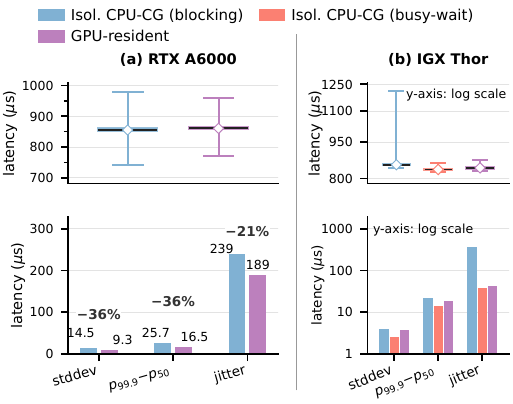}
\caption{Matrix multiplication under host-side contention:
latency distributions (top) and variation metrics (bottom).}
\label{fig:matmul-contention}
\end{figure}

\begin{figure}[t]
\centering
  \includegraphics[width=0.95\columnwidth]{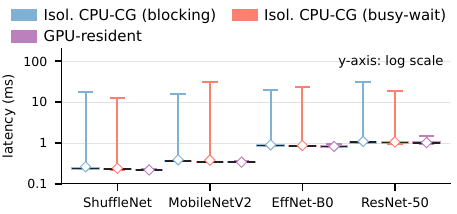}
\caption{Single-stage inference latency under host-side background
workloads on IGX Thor.}
\label{fig:contention-thor-inference}
\end{figure}

\Cref{fig:matmul-contention}a shows the latency distribution and variation metrics.
Compared with \textsf{CPU-CG}, GPU-resident reduces standard deviation from
\SI{14.5}{\micro\second} to \SI{9.3}{\micro\second},
$p_{99.9}-p_{50}$ from \SI{25.7}{\micro\second} to
\SI{16.5}{\micro\second}, and jitter from \SI{239}{\micro\second} to
\SI{189}{\micro\second}.
The improvements are not as large as the inference experiments, but \sysname
achieves this without reserving any host cores. Next, we examine whether the
benefits of \textsf{GPU-resident} carry over to other platforms.

\textbf{IGX Thor.}
We repeat the above two contention experiments on IGX~Thor to
test whether the result carries to an 
SoC platform. We also
add a CPU-side mitigation with \textsf{CPU-CG} that
busy-waits for CUDA completion (\texttt{cudaDeviceScheduleSpin}) instead
of using default blocking semantics. Busy-waiting can reduce wakeup latency, but
increases CPU utilization and can starve lower-priority work.

For the small emulated-I/O workload, \cref{fig:matmul-contention}b shows that
busy-waiting can make \textsf{CPU-CG} behave similarly to
\textsf{GPU-resident}. Still, CPU remains busy polling for events,
whereas CPU core counts are
limited in SoCs (see~\cref{fig:core_scaling}) and do not scale with more sensors,
requiring more CPU cores to run the polling threads.
For the inference workloads, however, \cref{fig:contention-thor-inference}
shows that busy-waiting does not remove the
costly CUDA calls, runtime paths, and other host overheads, so the
\textsf{CPU-CG} latency tail remains exposed to host contention.
Together, these results show why a mitigation in CPU-driven execution 
is not a robust substitute:
it can tune around one
surface of the shared CPU--GPU stack, but the effective solution depends on the
application, workload, and other factors. Indeed, the mitigations
aim to practically avoid the unpredictable components of the surface,
but do not address
them fundamentally.
The GPU-resident execution paradigm prevents any host-side
interference by design,
with a device-resident steady-state real-time path.

\subsubsection{Scaling to Concurrent Loops}\label{sec:eval-scaleup}

The next experiment tests the scalability of \sysname against concurrent
sensor processing loops,
running
matrix multiplications kernels.
We co-run 1, 5, 10, 15, and 20 such loops, each assigned a disjoint
4-SM static CUDA MPS partition~\cite{nvidia-mps-static-sm}, the smallest
partition on the A6000. As before, each \textsf{CPU-CG} loop is
also pinned to an
isolated CPU core. We vary the isolated-core count to 2/6/10, a realistic
limit for embedded SoCs where remaining cores serve the OS and
other system tasks. 
To demonstrate timing
predictability, scaling to concurrent loops must keep
the latency of individual loops stable.
We use the same measurement regime as in the emulated I/O experiment.
\begin{figure}[t]
  \centering
  \includegraphics[width=\columnwidth]{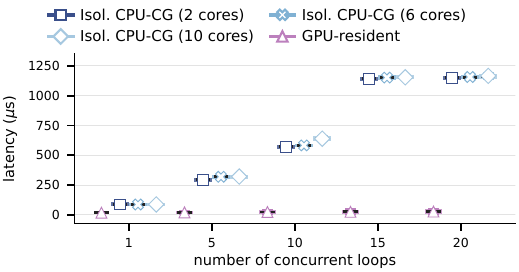}
  \caption{Concurrent loop with MPS static partitioning.}
  \label{fig:scaleup-boxplots}
\end{figure}
\begin{figure}[t]
  \centering
  \includegraphics[width=\columnwidth]{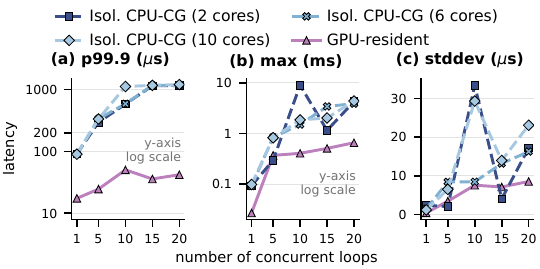}
  \caption{Concurrent loop latency tail and variation.}
  \label{fig:scaleup}
\end{figure}
\begin{figure}[t]
  \centering
  \includegraphics[width=\columnwidth]{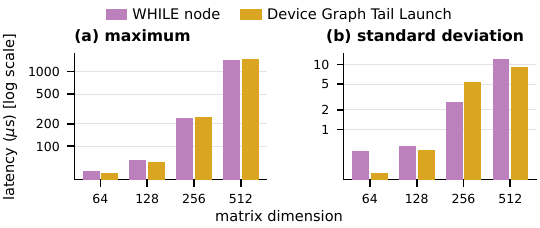}
  \caption{\textsc{WHILE} node vs. device graph tail launch.}
  \label{fig:whl-vs-dvg}
\end{figure}

\Cref{fig:scaleup-boxplots} shows the per-iteration latency distributions
for each measurement using Tukey box plots. Their whiskers reach the furthest
observations within $1.5\times$ the interquartile range from the box, and
outliers are omitted.
The figure shows that GPU-resident latency remains stable as the concurrent-loop
count increases, whereas no \textsf{CPU-CG} configuration maintains this stability.
\Cref{fig:scaleup} reveals the
p99.9 and max latencies (both log scale),
and standard deviation.
For \sysname, all three metrics rise from one to five loops, meaning
that device-side interference does not
disappear in \sysname.
Cache, memory-system, and other device-side effects
still need better hardware controls~\cite{jain19fractional}.
However, after this initial rise, the tail metrics remain relatively
stable: from 5 to 20 loops, \sysname's $p_{99.9}$ and maximum
latency stay within $2\times$, and stddev stays within
$2.5\times$. \textsf{CPU-CG} compounds those same
device-side effects with host-side work-queue scheduling and runlist
submission~\cite{bakita25compute}; its max
latency and stddev spike across loop and core counts.
Relative to its own five-loop values, \textsf{CPU-CG} worst-case latency grows
up to $30\times$,
$p_{99.9}$ up to $4\times$,
and stddev up to $16\times$.
At 10 concurrent loops, compared with the \textsf{CPU-CG} configuration using 10 isolated cores,
\sysname reserves no CPU cores and still reduces
standard deviation from \SI{29.3}{\micro\second} to
\SI{7.6}{\micro\second}, $p_{99.9}$ from \SI{1113.6}{\micro\second} to
\SI{50.2}{\micro\second}, and maximum latency from
\SI{1850.3}{\micro\second} to \SI{404.5}{\micro\second}. This shows that
\sysname makes concurrent loops more stable even when each \textsf{CPU-CG} loop
receives an isolated core.
Thus, \textsf{GPU-resident} demonstrates a more scalable and predictable 
design foundation for
multi-sensor processing applications with concurrent GPU-resident loops.

\subsubsection{Framework Overheads}\label{sec:eval-overhead}

The final experiments focus on the framework costs.
First, we show that the two GPU-resident iteration driver mechanisms,
\textsc{WHILE} node (\textsc{WHL}) and
Device Graph Tail Launch (\textsc{DVG}), have similar performance.
Then, we measure the \textsc{WHL} mechanism overheads.

\textsc{WHL} uses
CUDA Graph \textsc{while} conditional nodes, whereas \textsc{DVG}
achieves the same behavior by tail-launching device graphs.
\cref{fig:whl-vs-dvg} shows that the two mechanisms are equivalent on
a workload of 10 matrix multiplication operators.
Across matrix dimensions, max latency differs by at most
6\% and standard deviation by at most
$\SI{3}{\micro\second}$, with neither mechanism consistently ahead.
We support both mechanisms but make \textsc{WHL} the default because 
\textsc{DVG} does not support advanced CUDA Graph features such as conditional
nodes, which many CUDA-based libraries increasingly use~\cite{pytorch-cuda-graph-cond}.

We characterize the
overhead of \sysname on RTX A6000 with an 
empty processing kernel experiment in~\Cref{tab:overhead-breakdown}. 
These platform-specific fixed costs can be measured once per new platform
for timing analysis.

\begin{table}[!t]
\centering
\caption{Framework Overheads Breakdown}
\label{tab:overhead-breakdown}
\setlength{\tabcolsep}{2.5pt}
\begin{tabular}{@{}l|l|c@{}}
\textbf{Edge} & \textbf{Graph Component} & \textbf{Worst-case (\si{\micro\second})} \\
\hline
\textsc{drh} $\to$ \textsc{ctrl}     & handler-to-controller transition & 0.82 \\
\textsc{ctrl} kernel    & controller kernel & 1.67 \\
\textsc{ctrl} $\to$ \textsc{body}    & body-release transition & 3.69 \\
body $\to$ end     & body-completion transition & 1.37 \\
end $\to$ \textsc{drh}  & next-iteration transition & 5.67 \\
\hline
\end{tabular}
\end{table}

\subsection{Case Study: IMX274 Camera-to-display Pipeline}\label{sec:eval-imx274}

We evaluate the IMX274 camera-to-display application
from \cref{sec:ref-imx274} on IGX~Orin~\cite{nvidia-igx-orin} with an
RTX~6000~Ada.
The connected display is a
240~Hz G-SYNC monitor. This setup is representative of
industry-grade real-time streaming sensor
pipelines in medical devices~\cite{stryker-1688-aim},
surgical robotics~\cite{intuitive-da-vinci,orsi-holoscan-ar-surgery}, 
medical imaging and endoscopy~\cite{olympus-eus,jnj-monarch},
and broadly, in industrial automation domains.

We measure the performance of three configurations in this case study.
1) \textsf{CPU-driven (no GDR)} 
(GDR means GPUDirect~RDMA~\cite{nvidia-gpudirect-rdma} which is NVIDIA's
implementation of GPU-memory~RDMA) 
keeps both the data and control path on Linux on the CPU;
2) \textsf{CPU-driven (w/ GDR)} uses
GDR for the frame data path but still
uses CPU-side receive completion, frame selection, CUDA launch, and
display handoff. Vast majority of industry and
research deployments use these two mechanisms~\cite{olympus-eus,
orsi-holoscan-ar-surgery,barco-nexxis,deltacast-gpudirect-rdma}.
Both CPU-driven baselines are developed with
the CPU-driven Holoscan SDK~\cite{holoscan-sdk}.
3) \textsf{GPU-resident} is our proposed framework with GDR
and DOCA GPUNetIO for GPU-driven I/O. 
The HSB metadata handler has similar latency but relies on CPU-side polling,
making it vulnerable to host contention; we omit it since the gains are clear
even without added contention.

\begin{figure}[t]
  \centering
  \includegraphics[width=\columnwidth]{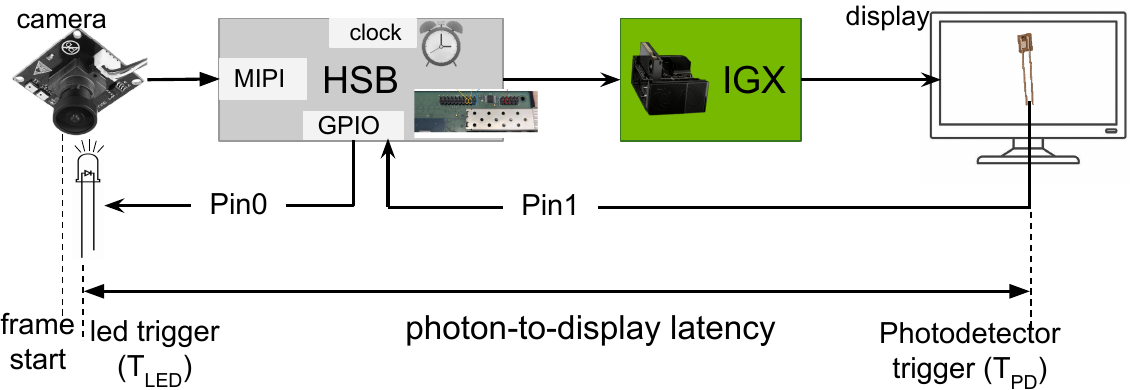}
  \caption{HSB-native photon-to-display latency tool.}
  \label{fig:hsb-native-p2d-tool}
\end{figure}

\textbf{Measurement.} We measure 1080p photon-to-display latency with a
HSB-native closed-loop tool, shown in~\cref{fig:hsb-native-p2d-tool}. The camera
stream enters the HSB over MIPI, and the corresponding frame payload is
forwarded over the sensor bridge and ConnectX-7 NIC to the IGX. The
configurations receive frames through either CPU memory (config~1)
or GPUDirect (config~2 and~3). Independently
of the data path, the HSB FPGA observes the IMX274 frame-start event
directly. For this rolling-shutter sensor, a HSB event sequencer drives
GPIO~Pin0 high at frame start, illuminating an
LED in the camera's field of view during the current frame readout. After
the frame is sent, processed, and displayed on the monitor, a photodiode mounted on
the display drives GPIO~Pin1 when the illuminated region appears. The
LED-trigger timestamp ($T_{\mathrm{LED}}$) and
photodiode timestamp ($T_{\mathrm{PD}}$) are latched by
the same FPGA clock, so the reported \emph{photon-to-display latency} 
is the hardware-timestamped
interval $T_{\mathrm{PD}} - T_{\mathrm{LED}}$.

\begin{figure}[t]
  \centering
  \includegraphics[width=\columnwidth]{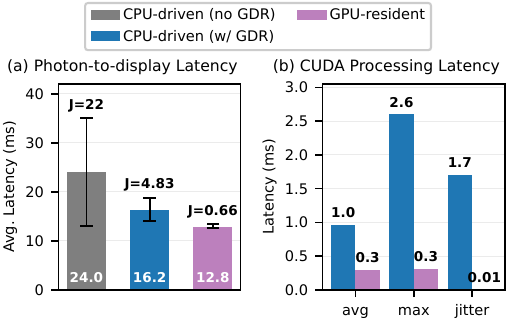}
  \caption{IMX274 camera-to-display case-study latency.}
  \label{fig:imx274-latency}
\end{figure}

\textbf{Result.} The left panel of \Cref{fig:imx274-latency}
compares photon-to-display latency (average with min--max whiskers).
The CPU-only Linux receiver has an average of
$\SI{24}{\milli\second}$ and a $\SI{22}{\milli\second}$ jitter.
Keeping the frame data path in GPU memory with GDR reduces
that span to $\SI{4.83}{\milli\second}$, but it is still affected by host
orchestration variation, which is problematic in surgical robotics and medical devices.
The GPU-resident graph eliminates the source of variance,
achieving sub-millisecond jitter
($\SI{0.66}{\milli\second}$), a $7.3\times$ reduction relative to
\textsf{CPU-driven (w/ GDR)} and
$33\times$ relative to \textsf{CPU-driven (no GDR)}.
The mean also moves below to
$\SI{12.81}{\milli\second}$ for \textsf{GPU-resident} versus
$\SI{16.2}{\milli\second}$ for \textsf{CPU-driven (w/ GDR)} and
$\SI{24}{\milli\second}$ for \textsf{CPU-driven (no GDR)}.
Crucially for time-sensitive domains, the
worst-case photon-to-display latency drops to
$\SI{13.35}{\milli\second}$ for \textsf{GPU-resident}, compared with
$\SI{18.83}{\milli\second}$ for \textsf{CPU-driven (w/ GDR)} and
$\SI{35}{\milli\second}$ for \textsf{CPU-driven (no GDR)}.
The right panel of \Cref{fig:imx274-latency} isolates the software
processing interval after frame arrival into the platform.
The \textsf{CPU-driven (w/ GDR)} pipeline averages
$\SI{0.96}{\milli\second}$, reaches $\SI{2.60}{\milli\second}$ at the
maximum, and has $J$=$\SI{1.70}{\milli\second}$. The GPU-resident path
averages $\SI{0.30}{\milli\second}$, reaches only
$\SI{0.31}{\milli\second}$ at the maximum, and has
$J$=$\SI{0.01}{\milli\second}$. The processing path is therefore
$3.2\times$ faster on average, $8.4\times$ lower at the maximum, and
$170\times$ lower in absolute jitter. Overall,
GPU-resident control removes the millisecond-scale latency tails
with host-driven execution for this streaming camera application.

\section{Discussion}\label{sec:disc}
\textbf{Scope.}
The primary significance of the results in~\cref{sec:eval}
is to demonstrate the structural benefits of \sysname, not just the quantitative improvements. A
CPU-driven loop, even if tuned with CUDA Graphs, GPUDirect~RDMA,
MPS, and core isolation, is susceptible to host latency tails.
Mitigations such as core isolation try to bypass the host-side challenges that
we identified, while \sysname removes the host altogether from steady-state.
Removing this source of variance produces the orders-of-magnitude reductions
in latency jitter observed in the streaming camera application.

For predictable latency,
\sysname mandates static device-memory layout, fixed clocks, no steady-state CPU
interaction, and an execution working set that fits in device memory.
These properties already match the common case for
certifiable sensor pipelines whose memory footprint and rate are known
at design time. The configurable data-not-ready sleep interval trades
polling latency for GPU utilization and power.
Furthermore, concurrent loops must be confined to a configured SM
partition per GPU-resident graph application.
Unmanaged compute- or
graphics-context coexistence can affect predictability.
For this reason, our case study
uses a GPU-resident display operator, avoiding a graphics context.
In our design, the CPU supervises the non-critical plane, while
the GPU runs the time-critical sensor processing loop.
Broadly, our
deliberately straightforward device-resident design
demonstrates empirically verified timing
predictability on industry-grade applications that no prior
work has achieved.

\textbf{GPU-resident graphs as asynchronous tasks.}
Each GPU-resident graph behaves as an asynchronous task that runs on a
fixed SM budget for the application lifetime. This work considers
independent tasks, including the concurrent loops evaluated on disjoint
SM partitions. Within one graph, the current handler abstraction
supports a sensor fusion pipeline by polling multiple sensors with parallel
CUDA threads before issuing a combined \texttt{dataReady}
signal, although our evaluation currently uses one sensor per handler.
Furthermore, the
framework does not define communication or synchronization among
distinct GPU-resident tasks or between GPU-resident and CPU tasks.
Designing and analyzing predictable communication among such
asynchronous tasks remains future work.

\vspace{6pt}
\section{Related Work}\label{sec:rw}
\vspace{4pt}

The sensor-I/O background and GPU mechanisms behind this work are already
covered in~\cref{sec:bg}. We now place our work in the context of the
closest adjacent research.

Persistent kernels~\cite{gupta12persistent} and graphs~\cite{cetre25persistentgraphs},
Whippletree~\cite{steinberger14whippletree},
PERKS~\cite{zhang22perks}, and
CharminG~\cite{choi21charming}
keep execution resident on the GPU and schedule work on device.
Their
mechanisms range from task queues and GPU-resident runtimes to CUDA
device graph launch~\cite{nvidia-device-graph-launch}.
Persistent kernels require modification to CUDA kernels, 
which GPU-resident does not need.
These systems share our goal
of CPU bypass, but none integrates sensing, actuation, 
and display I/O as first-class primitives.
In addition, most make dynamic on-device scheduling part of the
application runtime, affecting predictability and analysis.

Prior real-time GPU work spans driver-level schedulers and
managers~\cite{kato11timegraph,elliott13gpusync,capodieci18deadline,zou23rtgpu,han22reef,wang24gcaps},
middleware and pipeline schedulers for accelerator-backed sensor
workloads~\cite{enright24paam,xiang19dart,kang21lalarand,wang21sbeet,kang24rtswap}.
These systems improve predictability by regulating launches from the host, GPU
context arbitration, or task release in the host pipeline. 
In contrast, \sysname introduces a new paradigm that stops treating the GPU as a
host-subordinate client and 
avoids host-mediated orchestration in the time-critical path. 
It establishes the GPU as an independent real-time
accelerator~\cite{bakita25advantage} by moving 
the end-to-end path, including the I/O, onto the device. 
GPU resource partitioning and reservation 
mechanisms~\cite{nvidia-mps-static-sm,jain19fractional,bakita23compute,nvidia-green-context}
support this shift by exposing
dedicated SM partitions for concurrent timing-predictable workloads. Scheduling
a taskset therefore 
becomes a design-time partitioning problem: for each GPU-resident loop,
the response-time bounds the captured graph body on its
partition; for multiple loops, a
partition configuration must satisfy every loop's timing requirement.

\vspace{6pt}
\section{Conclusion and Future Work}\label{sec:conc}
\vspace{4pt}
Real-time ML-driven sensor processing needs both predictable
end-to-end latency and the advanced software stacks available on
Linux-hosted GPU platforms. This paper presented GPU-resident execution,
which keeps a steady-state closed-loop system
on the device for its lifetime, with GPU-native ingress, egress,
and control. By combining an operator and application API, device-side
data-ready handlers, GPU-visible I/O, and a long-lived CUDA Graph, the
design removes per-frame host wakeups and CUDA launches
from the critical path. The design yields a predictable system design
with device-side execution and
graph-transition costs, rather than Linux kernel, drivers, or CUDA
driver-dispatch latency tails. Across microbenchmarks and an industry-grade
camera-to-display application, experiments show that this
architectural shift tightens latency distributions and
reduces end-to-end jitter by orders of magnitude on NVIDIA SoCs and GPUs.

Future work will focus on sense--think--act loops~\cite{siegel03sense}
with broader GPU-driven I/O support, including HSB-style FPGA-backed
adapters for CAN/SPI-connected motor control
and direct control of SPI, GPIO, and I2C.
This work lays the groundwork for deploying
GPU-native action models such as GR00T~N1~\cite{nvidia25groot} inside
real-time control loops. As the broader trend in generative AI and robotics
shifts from producing pixels~\cite{huang23everypixel} to
producing actions~\cite{nvidia25groot}, our architecture
provides a timing-predictable systems foundation for GPU-native
sensor processing systems.

\balance

\bibliographystyle{IEEEtran}
\bibliography{references}

\begin{thebibliography}{100}
\providecommand{\url}[1]{#1}
\csname url@samestyle\endcsname
\providecommand{\newblock}{\relax}
\providecommand{\bibinfo}[2]{#2}
\providecommand{\BIBentrySTDinterwordspacing}{\spaceskip=0pt\relax}
\providecommand{\BIBentryALTinterwordstretchfactor}{4}
\providecommand{\BIBentryALTinterwordspacing}{\spaceskip=\fontdimen2\font plus
\BIBentryALTinterwordstretchfactor\fontdimen3\font minus
  \fontdimen4\font\relax}
\providecommand{\BIBforeignlanguage}[2]{{\expandafter\ifx\csname l@#1\endcsname\relax
\typeout{** WARNING: IEEEtran.bst: No hyphenation pattern has been}\typeout{** loaded for the language `#1'. Using the pattern for}\typeout{** the default language instead.}\else
\language=\csname l@#1\endcsname
\fi
#2}}
\providecommand{\BIBdecl}{\relax}
\BIBdecl

\bibitem{chen24endtoendad}
L.~Chen, P.~Wu, K.~Chitta, B.~Jaeger, A.~Geiger, and H.~Li, ``End-to-end
  autonomous driving: Challenges and frontiers,'' \emph{IEEE Transactions on
  Pattern Analysis and Machine Intelligence}, vol.~46, no.~12, pp.
  10\,164--10\,183, 2024.

\bibitem{roychoudhury23humanoid}
A.~Roychoudhury, S.~Khorshidi, S.~Agrawal \emph{et~al.}, ``Perception for
  humanoid robots,'' \emph{Current Robotics Reports}, vol.~4, pp. 127--140,
  2023.

\bibitem{lee22rtbev}
K.~Lee, H.-S. Cho, and J.~Lee, ``{RT-BEV}: An open-source framework for
  real-time {BEV} perception,'' in \emph{IEEE Real-Time Systems Symposium
  (RTSS)}, 2023.

\bibitem{nvidia-mb-drive}
{NVIDIA Corporation}, ``{Mercedes-Benz} and {NVIDIA} to build software-defined
  architecture for automated driving,'' NVIDIA Newsroom, 2020.

\bibitem{nvidia-hyundai-drive}
\BIBentryALTinterwordspacing
------, ``Hyundai motor, kia and {NVIDIA} expand strategic partnership for
  next-generation autonomous driving technology,'' NVIDIA Newsroom, 2025.
  [Online]. Available:
  \url{https://nvidianews.nvidia.com/news/hyundai-motor-kia-autonomous-driving}
\BIBentrySTDinterwordspacing

\bibitem{tijtgat17uav}
N.~Tijtgat, W.~Van~Ranst, B.~Volckaert, T.~Goedem{\'e}, and F.~De~Turck,
  ``Embedded real-time object detection for a {UAV} warning system,'' in
  \emph{IEEE International Conference on Computer Vision Workshops (ICCVW)},
  2017, pp. 2110--2118.

\bibitem{hossain19aerial}
S.~Hossain and D.-J. Lee, ``Deep learning-based real-time multiple-object
  detection and tracking from aerial imagery via a flying robot with
  {GPU}-based embedded devices,'' \emph{Sensors}, vol.~19, no.~15, 2019.

\bibitem{bostondynamics-spot-coreio}
{Boston Dynamics}, ``{Spot Core I/O},''
  \url{https://bostondynamics.com/products/payload/spot-core-io/}, 2025.

\bibitem{nvidia-holoscan-medical}
{NVIDIA Corporation}, ``{AI} platform to bring real-time sensing to medical
  devices,''
  \url{https://blogs.nvidia.com/blog/clara-holoscan-real-time-ai-medical-devices/},
  2022.

\bibitem{ahmed25surgical}
F.~A. Ahmed, M.~Yousef, M.~A. Ahmed \emph{et~al.}, ``Deep learning for surgical
  instrument recognition and segmentation in robotic-assisted surgeries: A
  systematic review,'' \emph{Artificial Intelligence Review}, vol.~58, 2025.

\bibitem{nvidia-visual-inspection}
{NVIDIA Corporation}, ``Build a real-time visual inspection pipeline with
  {NVIDIA} {TAO} and {NVIDIA} {DeepStream},'' NVIDIA Technical Blog, 2025.

\bibitem{liu23surface}
Y.~Liu, C.~Zhang, and X.~Dong, ``A survey of real-time surface defect
  inspection methods based on deep learning,'' \emph{Artificial Intelligence
  Review}, vol.~56, pp. 12\,131--12\,170, 2023.

\bibitem{nvidia-jetson-modules}
{NVIDIA Corporation}, ``{Jetson} modules, support, ecosystem, and lineup,''
  \url{https://developer.nvidia.com/embedded/jetson-modules}, 2025.

\bibitem{nvidia-igx-orin}
------, ``{NVIDIA} {IGX} {Orin},''
  \url{https://www.nvidia.com/en-us/edge-computing/products/igx/}, 2026.

\bibitem{nvidia-igx-thor}
------, ``{NVIDIA} {IGX} {Thor},''
  \url{https://www.nvidia.com/en-us/edge-computing/products/igx/}, 2026.

\bibitem{amd-zynq-ultrascale-mpsoc}
{Advanced Micro Devices, Inc.}, ``{AMD} {Zynq} {UltraScale+} {MPSoC},''
  \url{https://www.amd.com/en/products/adaptive-socs-and-fpgas/soc/zynq-ultrascale-plus-mpsoc.html},
  2026, eG and EV variants include Arm Mali-400 MP2 GPU.

\bibitem{nvidia-cuda-graphs}
{NVIDIA Corporation}, ``Getting started with {CUDA} graphs,''
  \url{https://developer.nvidia.com/blog/cuda-graphs/}, 2019.

\bibitem{de2022operating}
D.~B. de~Oliveira, D.~Casini, and T.~Cucinotta, ``Operating system noise in the
  linux kernel,'' \emph{IEEE Transactions on Computers}, vol.~72, no.~1, pp.
  196--207, 2022.

\bibitem{bristot20linux}
D.~B. de~Oliveira, D.~Casini, R.~S. de~Oliveira, and T.~Cucinotta,
  ``Demystifying the real-time {Linux} scheduling latency,'' in \emph{Euromicro
  Conference on Real-Time Systems (ECRTS)}, 2020.

\bibitem{zhang06interrupt}
Y.~Zhang and R.~West, ``Process-aware interrupt scheduling and accounting,'' in
  \emph{IEEE Real-Time Systems Symposium (RTSS)}, 2006.

\bibitem{madden19linuxrt}
M.~M. Madden, ``Challenges using {Linux} as a real-time operating system,'' in
  \emph{AIAA SciTech Forum}, 2019, {NASA} Technical Reports Server, Document ID
  20200002390.

\bibitem{missimer16io}
E.~Missimer, K.~Missimer, and R.~West, ``Mixed-criticality scheduling with
  {I/O},'' in \emph{Euromicro Conference on Real-Time Systems (ECRTS)}, 2016.

\bibitem{linux-rt-theory}
{Linux Kernel Documentation}, ``Real-time linux: Theory of operation,''
  \url{https://docs.kernel.org/core-api/real-time/theory.html}, 2026.

\bibitem{reghenzani19preempt}
F.~Reghenzani, G.~Massari, and W.~Fornaciari, ``The real-time linux kernel: A
  survey on {PREEMPT\_RT},'' \emph{ACM Computing Surveys}, vol.~52, no.~1,
  2019.

\bibitem{voneicken95unet}
T.~von Eicken, A.~Basu, V.~Buch, and W.~Vogels, ``{U-Net}: A user-level network
  interface for parallel and distributed computing,'' in \emph{ACM Symposium on
  Operating Systems Principles (SOSP)}, 1995, pp. 40--53.

\bibitem{dpdk}
{DPDK Project}, ``{Data Plane Development Kit} ({DPDK}),''
  \url{https://www.dpdk.org/}, 2026.

\bibitem{nvidia-gpudirect-rdma}
{NVIDIA Corporation}, ``{GPUDirect} {RDMA} documentation,''
  \url{https://docs.nvidia.com/cuda/gpudirect-rdma/}, 2024.

\bibitem{amd-rocm-rdma}
{Advanced Micro Devices, Inc.}, ``{ROCm} gpu-enabled message passing
  interface,''
  \url{https://rocm.docs.amd.com/en/develop/how-to/gpu-enabled-mpi.html}, 2026.

\bibitem{nvidia-gpunetio}
{NVIDIA Corporation}, ``{DOCA} {GPUNetIO}: Real-time {GPU}-driven network
  processing,''
  \url{https://docs.nvidia.com/doca/sdk/DOCA-GPUNetIO/index.html}, 2026.

\bibitem{amd-rocm-xio}
{Advanced Micro Devices, Inc.}, ``{rocm-xio}: A {ROCm} library for
  {GPU}-initiated {IO},'' \url{https://github.com/ROCm/rocm-xio}, 2026.

\bibitem{betts13wcetgpu}
A.~Betts and A.~F. Donaldson, ``Estimating the {WCET} of {GPU}-accelerated
  applications using hybrid analysis,'' in \emph{Euromicro Conference on
  Real-Time Systems (ECRTS)}, 2013, pp. 193--202.

\bibitem{berezovskyi14cuda}
K.~Berezovskyi, L.~Santinelli, K.~Bletsas, and E.~Tovar, ``{WCET}
  measurement-based and extreme value theory characterisation of {CUDA}
  kernels,'' in \emph{International Conference on Real-Time Networks and
  Systems (RTNS)}, 2014, pp. 279--288.

\bibitem{jeanmougin25wcetgpu}
L.~Jeanmougin, T.~Carle, and C.~Rochange, ``Bounding the {WCET} of a {GPU}
  thread block with a multi-phase representation of warps execution,'' in
  \emph{Euromicro Conference on Real-Time Systems (ECRTS)}, 2025, pp.
  11:1--11:26.

\bibitem{li12gklee}
G.~Li, P.~Li, G.~Sawaya, G.~Gopalakrishnan, I.~Ghosh, and S.~P. Rajan,
  ``{GKLEE}: Concolic verification and test generation for {GPU}s,'' in
  \emph{ACM SIGPLAN Symposium on Principles and Practice of Parallel
  Programming (PPoPP)}, 2012, pp. 215--224.

\bibitem{betts12gpuverify}
A.~Betts, N.~Chong, A.~F. Donaldson, S.~Qadeer, and P.~Thomson, ``{GPUVerify}:
  A verifier for {GPU} kernels,'' in \emph{ACM SIGPLAN Conference on
  Object-Oriented Programming, Systems, Languages, and Applications (OOPSLA)},
  2012, pp. 113--132.

\bibitem{holoscan-sdk}
{NVIDIA Corporation}, ``{Holoscan} {SDK}: {AI} sensor processing platform,''
  \url{https://docs.nvidia.com/holoscan/sdk-user-guide/}, 2026.

\bibitem{ros2-foundation}
{Open Source Robotics Foundation}, ``{ROS}~2 documentation,''
  \url{https://docs.ros.org/}, 2024.

\bibitem{nvidia-deepstream}
{NVIDIA Corporation}, ``{NVIDIA} {DeepStream} {SDK},''
  \url{https://developer.nvidia.com/deepstream-sdk}, 2026.

\bibitem{nvidia-isaac-ros}
------, ``{NVIDIA} {Isaac} {ROS},''
  \url{https://developer.nvidia.com/isaac-ros}, 2026.

\bibitem{amd-vvas}
{Advanced Micro Devices, Inc.}, ``{Vitis} video analytics {SDK} ({VVAS}),''
  \url{https://www.amd.com/en/developer/resources/vitis-video-analytics-sdk.html},
  2026.

\bibitem{danish11questvcpu}
M.~Danish, Y.~Li, and R.~West, ``Virtual-{CPU} scheduling in the {Quest}
  operating system,'' in \emph{IEEE Real-Time and Embedded Technology and
  Applications Symposium (RTAS)}, 2011, pp. 169--179.

\bibitem{qnx-neutrino}
{BlackBerry QNX}, ``{QNX} {Neutrino} {RTOS}: System architecture guide,''
  \url{https://www.qnx.com/developers/docs/7.1/com.qnx.doc.neutrino.sys_arch/topic/intro.html},
  2024.

\bibitem{li14questv}
Y.~Li, E.~Missimer, and R.~West, ``Predictable migration and communication in
  the {Quest-V} multikernel,'' arXiv:1310.6301, 2014.

\bibitem{sinha21driveos}
S.~Sinha and R.~West, ``Towards an integrated vehicle management system in
  {DriveOS},'' \emph{ACM Transactions on Embedded Computing Systems}, vol.~20,
  no.~5s, 2021.

\bibitem{sinha22thesis}
S.~Sinha, ``Towards a centralized multicore automotive system,'' Ph.D.
  dissertation, Boston University, 2022. [Online]. Available:
  \url{https://hdl.handle.net/2144/45640}

\bibitem{nvidia-drive-os}
{NVIDIA Corporation}, ``{DRIVE} {OS} {QNX} installation guide,''
  \url{https://developer.nvidia.com/docs/drive/drive-os/6.0.6/public/drive-os-qnx-installation/index.html},
  2024.

\bibitem{nvidia-drive-qnx-cuda}
------, ``{CUDA} safety installation guide for {DRIVE} {OS} ({QNX}),''
  \url{https://docs.nvidia.com/drive/drive-os-5.2.0.0L/drive-qsg-nv/cuda-safety-installation-guide/},
  2024.

\bibitem{olmedo20cuda}
I.~S. Olmedo, N.~Capodieci, J.~L. Mart{\'{\i}}nez, A.~Marongiu, and
  M.~Bertogna, ``Dissecting the {CUDA} scheduling hierarchy: A performance and
  predictability perspective,'' in \emph{IEEE Real-Time and Embedded Technology
  and Applications Symposium (RTAS)}, 2020.

\bibitem{bakita24demystifying}
J.~Bakita and J.~H. Anderson, ``Demystifying {NVIDIA} {GPU} internals to enable
  reliable {GPU} management,'' in \emph{IEEE Real-Time and Embedded Technology
  and Applications Symposium (RTAS)}, 2024.

\bibitem{liu24autonomy}
S.~Liu, R.~Wagle, J.~H. Anderson, M.~Yang, C.~Zhang, and Y.~Li, ``Autonomy
  today: Many delay-prone black boxes,'' in \emph{36th Euromicro Conference on
  Real-Time Systems (ECRTS 2024)}, ser. Leibniz International Proceedings in
  Informatics (LIPIcs), vol. 298.\hskip 1em plus 0.5em minus 0.4em\relax
  Schloss Dagstuhl -- Leibniz-Zentrum f{\"u}r Informatik, 2024, pp.
  12:1--12:27.

\bibitem{nvidia-gpudirect-benchmark}
{NVIDIA Corporation}, ``Benchmarking {GPUDirect} {RDMA} on modern server
  platforms,''
  \url{https://devblogs.nvidia.com/parallelforall/benchmarking-gpudirect-rdma-on-modern-server-platforms/},
  2017.

\bibitem{sapio21scaling}
A.~Sapio, M.~Canini, C.-Y. Ho, J.~Nelson, P.~Sudber \emph{et~al.}, ``Scaling
  distributed machine learning with in-network aggregation,'' in \emph{USENIX
  Symposium on Networked Systems Design and Implementation (NSDI)}, 2021.

\bibitem{holoscan-sensor-bridge}
{NVIDIA Corporation}, ``{Holoscan} {Sensor} {Bridge} documentation,''
  \url{https://docs.nvidia.com/holoscan/sensor-bridge/}, 2026.

\bibitem{holoscan-sensor-bridge-github}
------, ``{Holoscan} {Sensor} {Bridge} (open-source fpga {IP} and host
  software),'' \url{https://github.com/nvidia-holoscan/holoscan-sensor-bridge},
  2026.

\bibitem{agostini17offloading}
E.~Agostini, D.~Rossetti, and S.~Potluri, ``Offloading communication control
  logic in {GPU} accelerated applications,'' in \emph{IEEE/ACM Int. Symp. on
  Cluster, Cloud and Grid Computing (CCGrid)}, 2017, pp. 248--257.

\bibitem{agostini18gpudirectasync}
------, ``{GPUDirect Async}: Exploring {GPU} synchronous communication
  techniques for {InfiniBand} clusters,'' \emph{Journal of Parallel and
  Distributed Computing}, vol. 114, pp. 28--45, 2018.

\bibitem{nvidia-gpunetio-blog}
{NVIDIA Corporation}, ``Inline {GPU} packet processing with {NVIDIA} {DOCA}
  {GPUNetIO},''
  \url{https://developer.nvidia.com/blog/inline-gpu-packet-processing-with-nvidia-doca-gpunetio/},
  2023.

\bibitem{nvidia-gpunetio-rdma}
------, ``Unlocking {GPU}-accelerated {RDMA} with {NVIDIA} {DOCA} {GPUNetIO},''
  \url{https://developer.nvidia.com/blog/unlocking-gpu-accelerated-rdma-with-nvidia-doca-gpunetio/},
  2024.

\bibitem{liao24gpucentric}
R.~Liao \emph{et~al.}, ``The landscape of {GPU}-centric communication,''
  \emph{arXiv preprint arXiv:2409.09874}, 2024.

\bibitem{holoscan-cuda-stream-handling}
{NVIDIA Corporation}, ``{CUDA} stream handling in {Holoscan} applications,''
  \url{https://docs.nvidia.com/holoscan/sdk-user-guide/holoscan_cuda_stream_handling.html},
  2026.

\bibitem{nvidia-cuda-graphs-conditional}
------, ``Dynamic control flow in {CUDA} graphs with conditional nodes,''
  \url{https://developer.nvidia.com/blog/dynamic-control-flow-in-cuda-graphs-with-conditional-nodes/},
  2024.

\bibitem{nvidia-device-graph-launch}
S.~Stevenson, ``Enabling dynamic control flow in {CUDA} graphs with device
  graph launch,''
  \url{https://developer.nvidia.com/blog/enabling-dynamic-control-flow-in-cuda-graphs-with-device-graph-launch/},
  2022.

\bibitem{pytorch-cuda-graph-cond}
\BIBentryALTinterwordspacing
{PyTorch Contributors}, ``Add support for \texttt{torch.cond} with {CUDA}
  graphs,'' GitHub Pull Request \#168912, 2026. [Online]. Available:
  \url{https://github.com/pytorch/pytorch/pull/168912}
\BIBentrySTDinterwordspacing

\bibitem{holoscan-gpu-resident}
{NVIDIA Corporation}, ``{GPU}-resident graph execution api,''
  \url{https://docs.nvidia.com/holoscan/sdk-user-guide/gpu_resident.html},
  2026.

\bibitem{nvidia-mps-static-sm}
------, ``{CUDA} {Multi-Process Service}: Static {SM} partitioning,''
  \url{https://docs.nvidia.com/deploy/mps/}, 2026.

\bibitem{nvidia-smi}
------, ``{NVIDIA} system management interface,''
  \url{https://docs.nvidia.com/deploy/nvidia-smi/}, 2026.

\bibitem{jain19fractional}
S.~Jain, I.~Baek, S.~Wang, and R.~Rajkumar, ``Fractional {GPU}s: Software-based
  compute and memory bandwidth reservation for {GPU}s,'' in \emph{IEEE
  Real-Time and Embedded Technology and Applications Symposium (RTAS)}, 2019,
  pp. 29--41.

\bibitem{zhang25sgdrc}
Y.~Zhang, H.~Yu, C.~Han, C.~Wang, B.~Lu, Y.~Li, Z.~Jiang, Y.~Li, X.~Chu, and
  H.~Li, ``{SGDRC}: Software-defined dynamic resource control for concurrent
  {DNN} inference on {NVIDIA} {GPU}s,'' in \emph{Proceedings of the 30th ACM
  SIGPLAN Annual Symposium on Principles and Practice of Parallel Programming},
  2025, pp. 267--281.

\bibitem{intel-rdt}
{Intel Corporation}, ``{Intel} {Resource Director Technology} ({Intel} {RDT})
  on 2nd generation {Intel} {Xeon} scalable processors,''
  \url{https://www.intel.com/content/dam/develop/external/us/en/documents/180115-intel-rdtcascadelake-serverreferencemanual-806717.pdf},
  2019.

\bibitem{nvidia-npp}
{NVIDIA Corporation}, ``{NVIDIA} performance primitives ({NPP}),''
  \url{https://developer.nvidia.com/npp}, 2026.

\bibitem{leopard-imx274-mipi-cs}
{Leopard Imaging Inc.}, ``{LI-IMX274-MIPI-CS}: {IMX274} 8m 4k camera module
  with {CS} lens,''
  \url{https://leopardimaging.com/product/robotics-cameras/cis-2-mipi-modules/i-pex-mipi-camera-modules/rolling-shutter-mipi-cameras/8-49mp-imx274/li-imx274-mipi-cs/},
  2026.

\bibitem{ibta-spec}
{InfiniBand Trade Association}, ``{InfiniBand} architecture specification
  volume 1,'' InfiniBand Trade Association, Tech. Rep., 2015, release 1.3.
  Defines RDMA Write and RDMA Write with Immediate operations.

\bibitem{nvidia-rdma-aware-programming}
{NVIDIA Corporation}, ``{RDMA} aware networks programming user manual,''
  \url{https://docs.nvidia.com/rdma-aware-networks-programming-user-manual-1-7.pdf},
  2026.

\bibitem{nvidia-connectx7}
------, ``{ConnectX-7} {SmartNIC},''
  \url{https://www.nvidia.com/en-us/networking/ethernet-adapters/connectx-7/},
  2024.

\bibitem{linux-drm}
{Linux Kernel Documentation}, ``Direct rendering manager ({DRM}) internals,''
  \url{https://docs.kernel.org/gpu/drm-internals.html}, 2026.

\bibitem{sinha24deterministic}
S.~Sinha, S.~Dwivedi, and M.~Azizian, ``Towards deterministic end-to-end
  latency for medical {AI} systems in {NVIDIA} {Holoscan},'' in \emph{ACM/IEEE
  Int. Conf. on Cyber-Physical Systems (ICCPS)}, 2024, pp. 235--246.

\bibitem{holohub-imx274-gpu-resident}
{NVIDIA Corporation}, ``Source code for {IMX274} {GPU}-resident application,''
  \url{https://nvidia-holoscan.github.io/holohub/applications/imx274_gpu_resident/},
  2026.

\bibitem{buttazzo11hard}
G.~C. Buttazzo, \emph{Hard Real-Time Computing Systems: Predictable Scheduling
  Algorithms and Applications}, 3rd~ed.\hskip 1em plus 0.5em minus 0.4em\relax
  Springer, 2011.

\bibitem{cervin04jittermargin}
A.~Cervin, B.~Lincoln, J.~Eker, K.-E. {\AA}rz{\'e}n, and G.~C. Buttazzo, ``The
  jitter margin and its application in the design of real-time control
  systems,'' in \emph{International Conference on Real-Time and Embedded
  Computing Systems and Applications (RTCSA)}, 2004, pp. 1--9.

\bibitem{ma18shufflenetv2}
N.~Ma, X.~Zhang, H.-T. Zheng, and J.~Sun, ``{ShuffleNet V2}: Practical
  guidelines for efficient {CNN} architecture design,'' in \emph{European
  Conference on Computer Vision (ECCV)}, 2018, pp. 116--131.

\bibitem{sandler18mobilenetv2}
M.~Sandler, A.~Howard, M.~Zhu, A.~Zhmoginov, and L.-C. Chen, ``{MobileNetV2}:
  Inverted residuals and linear bottlenecks,'' in \emph{IEEE Conference on
  Computer Vision and Pattern Recognition (CVPR)}, 2018, pp. 4510--4520.

\bibitem{tan19efficientnet}
M.~Tan and Q.~V. Le, ``{EfficientNet}: Rethinking model scaling for
  convolutional neural networks,'' in \emph{International Conference on Machine
  Learning (ICML)}, 2019, pp. 6105--6114.

\bibitem{he16resnet}
K.~He, X.~Zhang, S.~Ren, and J.~Sun, ``Deep residual learning for image
  recognition,'' in \emph{IEEE Conference on Computer Vision and Pattern
  Recognition (CVPR)}, 2016, pp. 770--778.

\bibitem{simonyan15vgg}
K.~Simonyan and A.~Zisserman, ``Very deep convolutional networks for
  large-scale image recognition,'' in \emph{International Conference on
  Learning Representations (ICLR)}, 2015.

\bibitem{kosaian21arithmetic}
J.~Kosaian and K.~V. Rashmi, ``Arithmetic-intensity-guided fault tolerance for
  neural network inference on {GPU}s,'' in \emph{International Conference for
  High Performance Computing, Networking, Storage and Analysis (SC)}, 2021, pp.
  1--14.

\bibitem{stress-ng}
C.~I. King, ``\texttt{stress-ng}: Stress test a computer system,''
  \url{https://github.com/ColinIanKing/stress-ng}, 2026.

\bibitem{bakita25compute}
J.~Bakita and J.~H. Anderson, ``Hardware compute partitioning on {NVIDIA}
  {GPU}s for composable systems,'' in \emph{37th Euromicro Conference on
  Real-Time Systems (ECRTS)}, ser. Leibniz International Proceedings in
  Informatics (LIPIcs), vol. 335.\hskip 1em plus 0.5em minus 0.4em\relax
  Schloss Dagstuhl -- Leibniz-Zentrum f{\"u}r Informatik, 2025, pp.
  21:1--21:25.

\bibitem{stryker-1688-aim}
{Stryker Corporation}, ``{1688 Advanced Imaging Modalities} ({AIM}) {4K}
  platform,''
  \url{https://www.stryker.com/us/en/endoscopy/systems/1688-aim-platform.html},
  2026.

\bibitem{intuitive-da-vinci}
{Intuitive Surgical, Inc.}, ``{da Vinci} robotic surgical systems,''
  \url{https://www.intuitive.com/en-us/products-and-services/da-vinci}, 2026.

\bibitem{orsi-holoscan-ar-surgery}
{Orsi Academy}, ``Orsi academy brings real-time {AI} to the {OR} for a
  first-of-its-kind augmented reality surgery,''
  \url{https://www.orsi-online.com/news/}, 2023.

\bibitem{olympus-eus}
{Olympus Corporation}, ``Endoscopic ultrasound ({EUS}),''
  \url{https://medical.olympusamerica.com/procedure/endoscopic-ultrasound-eus},
  2026.

\bibitem{jnj-monarch}
{Johnson \& Johnson MedTech}, ``{MONARCH} platform for robotic-assisted
  bronchoscopy,''
  \url{https://www.jnjmedtech.com/en-US/products/robotics/monarch-platform/bronchoscopy/},
  2026.

\bibitem{barco-nexxis}
{Barco NV}, ``{Nexxis}: Uncompressed {AV}-over-{IP} platform,''
  \url{https://www.barco.com/en/product/nexxis}, 2026.

\bibitem{deltacast-gpudirect-rdma}
{DELTACAST}, ``{NVIDIA} {GPUDirect} {RDMA} for video capture and processing,''
  \url{https://www.deltacast.tv/technologies/gpudirect-rdma}, 2026.

\bibitem{gupta12persistent}
K.~Gupta, J.~A. Stuart, and J.~D. Owens, ``A study of persistent threads style
  {GPU} programming for {GPGPU} workloads,'' in \emph{Innovative Parallel
  Computing (InPar)}, 2012.

\bibitem{cetre25persistentgraphs}
C.~Cetre, F.~Ferreira, R.~Barrere, and D.~Gratadour, ``Introducing {GPU}
  persistent graphs for time-sensitive workflows,'' in \emph{ACM/SPEC
  International Conference on Performance Engineering (ICPE)}, 2025, pp.
  181--191.

\bibitem{steinberger14whippletree}
M.~Steinberger, M.~Kenzel, P.~Boechat, B.~Kerbl, M.~Dokter, and D.~Schmalstieg,
  ``{Whippletree}: Task-based scheduling of dynamic workloads on the {GPU},''
  \emph{ACM Trans. Graph. (Proc. SIGGRAPH Asia)}, vol.~33, no.~6, 2014.

\bibitem{zhang22perks}
L.~Zhang, M.~Wahib, P.~Chen, J.~Meng, X.~Wang, T.~Endo, and S.~Matsuoka,
  ``{PERKS}: A locality-optimized execution model for iterative memory-bound
  {GPU} applications,'' \emph{arXiv preprint arXiv:2204.02064}, 2022.

\bibitem{choi21charming}
J.~Choi, D.~F. Richards, and L.~V. Kale, ``{CharminG}: A scalable
  {GPU}-resident runtime system,'' in \emph{ACM International Symposium on
  High-Performance Parallel and Distributed Computing (HPDC)}, 2021, pp.
  261--262.

\bibitem{kato11timegraph}
S.~Kato, K.~Lakshmanan, R.~Rajkumar, and Y.~Ishikawa, ``{TimeGraph}: {GPU}
  scheduling for real-time multi-tasking environments,'' in \emph{USENIX Annual
  Technical Conference (USENIX ATC)}, 2011.

\bibitem{elliott13gpusync}
G.~A. Elliott, B.~C. Ward, and J.~H. Anderson, ``{GPUSync}: A framework for
  real-time {GPU} management,'' in \emph{IEEE Real-Time Systems Symposium
  (RTSS)}, 2013, pp. 33--44.

\bibitem{capodieci18deadline}
N.~Capodieci, R.~Cavicchioli, M.~Bertogna, and A.~Paramakuru, ``Deadline-based
  scheduling for {GPU} with preemption support,'' in \emph{IEEE Real-Time
  Systems Symposium (RTSS)}, 2018.

\bibitem{zou23rtgpu}
A.~Zou, J.~Li, C.~D. Gill, and X.~Zhang, ``{RTGPU}: Real-time {GPU} scheduling
  of hard deadline parallel tasks with fine-grain utilization,'' \emph{IEEE
  Trans. Parallel Distrib. Syst.}, 2023.

\bibitem{han22reef}
M.~Han, H.~Zhang, R.~Chen, and H.~Chen, ``Microsecond-scale preemption for
  concurrent {GPU}-accelerated {DNN} inferences,'' in \emph{USENIX Symposium on
  Operating Systems Design and Implementation (OSDI)}, 2022, pp. 539--558.

\bibitem{wang24gcaps}
Y.~Wang, C.~Liu, D.~Wong, and H.~Kim, ``{GCAPS}: {GPU} context-aware preemptive
  priority-based scheduling for real-time tasks,'' in \emph{Euromicro
  Conference on Real-Time Systems (ECRTS)}, 2024.

\bibitem{enright24paam}
D.~Enright, Y.~Xiang, H.~Choi, and H.~Kim, ``{PAAM}: A framework for
  coordinated and priority-driven accelerator management in {ROS}~2,'' in
  \emph{IEEE Real-Time and Embedded Technology and Applications Symposium
  (RTAS)}, 2024.

\bibitem{xiang19dart}
Y.~Xiang and H.~Kim, ``Pipelined data-parallel {CPU/GPU} scheduling for
  multi-{DNN} real-time inference,'' in \emph{IEEE Real-Time Systems Symposium
  (RTSS)}, 2019, pp. 392--405.

\bibitem{kang21lalarand}
W.~Kang, K.~Lee, J.~Lee, I.~Shin, and H.~S. Chwa, ``{LaLaRAND}: Flexible
  layer-by-layer {CPU/GPU} scheduling for real-time {DNN} tasks,'' in
  \emph{IEEE Real-Time Systems Symposium (RTSS)}, 2021.

\bibitem{wang21sbeet}
Y.~Wang, M.~Karimi, Y.~Kim, and H.~Kim, ``Balancing energy efficiency and
  real-time performance in {GPU} scheduling,'' in \emph{IEEE Real-Time Systems
  Symposium (RTSS)}, 2021.

\bibitem{kang24rtswap}
F.~Kang, K.~Lee, and J.~Lee, ``{RT-Swap}: Addressing {GPU} memory bottlenecks
  for real-time multi-{DNN} inference,'' in \emph{IEEE Real-Time and Embedded
  Technology and Applications Symposium (RTAS)}, 2024.

\bibitem{bakita25advantage}
J.~Bakita and J.~H. Anderson, ``The advantage of the {GPU} as a real-time {AI}
  accelerator,'' \emph{Real-Time Systems}, vol.~61, pp. 275--280, 2025.

\bibitem{bakita23compute}
------, ``Hardware compute partitioning on {NVIDIA} {GPU}s,'' in \emph{IEEE
  Real-Time and Embedded Technology and Applications Symposium (RTAS)}, 2023,
  pp. 54--66.

\bibitem{nvidia-green-context}
{NVIDIA Corporation}, ``{CUDA} {Green} context for spatial partitioning,''
  \url{https://developer.nvidia.com/blog/dynamic-load-balancing-in-cuda-with-green-contexts/},
  2024.

\bibitem{siegel03sense}
M.~Siegel, ``The sense-think-act paradigm revisited,'' in \emph{1st
  International Workshop on Robotic Sensing (ROSE)}, 2003, p.~5.

\bibitem{nvidia25groot}
{NVIDIA Corporation}, ``{GR00T N1}: An open foundation model for generalist
  humanoid robots,'' arXiv:2503.14734; \url{https://arxiv.org/abs/2503.14734},
  2025.

\bibitem{huang23everypixel}
J.~Huang, ``Every single pixel will be generated soon. not rendered:
  generated,'' Remarks at {NVIDIA} {GTC} 2023;
  \url{https://www.tomshardware.com/pc-components/gpus/rtx-off-ai-on-jensen-says-well-see-fully-ai-generated-games-in-5-10-years},
  2023.

\end{thebibliography}

\end{document}